\documentclass[conference]{IEEEtran}
\IEEEoverridecommandlockouts

\usepackage{cite}
\usepackage{amsmath,amssymb,amsfonts}
\usepackage{algorithmic}
\usepackage{graphicx}
\usepackage{textcomp}
\usepackage{xcolor}
\usepackage{amsthm}
\usepackage{multirow}
\usepackage{booktabs} 
\usepackage{setspace} 
\usepackage{threeparttable}  
\usepackage{makecell}        

\newtheoremstyle{asccstyle}{3pt}{3pt}{\normalfont}{\parindent}{\itshape}{.}{0.5em}{}
\theoremstyle{asccstyle}
\newtheorem{theorem}{Theorem}
\newtheorem{lemma}{Lemma}
\newtheorem{corollary}{Corollary}
\newtheorem{definition}{Definition}
\newtheorem{remark}{Remark}
\newtheorem{proposition}{Proposition}
\def\BibTeX{{\rm B\kern-.05em{\sc i\kern-.025em b}\kern-.08em
    T\kern-.1667em\lower.7ex\hbox{E}\kern-.125emX}}
\begin{document}

\title{Compositional Data-Driven Small-Signal Stability Assessment of Power Systems via Output Differential Passivity}

\author{
    \IEEEauthorblockN{Peng Yang, \IEEEmembership{Member, IEEE}, Xiaohui Zhang, Xi Ru, Feng Liu,  \IEEEmembership{Senior Member, IEEE}}
    \thanks{This work was supported by the National Natural Science Foundation of China (No. 52407137) and the Young Talent Fund of Association for Science and Technology in Shaanxi, China (No. 20250431). (Corresponding author: Peng Yang)}
    }

\maketitle

\begin{abstract}
This paper proposes a compositional data-driven small-signal stability assessment framework for interconnected power systems based on output differential passivity (ODP). Without requiring explicit device models, local ODP matrices and scalar indices are estimated directly from measurement trajectories. When internal states are available, trajectory-based linear matrix inequalities (LMIs) are constructed from input-state measurements to estimate the local ODP matrices and scalar indices. When internal states are unavailable, an extended-state representation is constructed from input-output measurements, enabling the same assessment using only external signals. The estimated ODP matrices are combined with the network coupling matrix for compositional assessment of power system stability, while their scalar indices enable decentralized assessment. To address the computational and numerical challenges of trajectory-based LMIs, we introduce right-inverse whitening (RIW) to remove redundant behavioral coordinates and improve computational efficiency, and apply instrumental-variable (IV) preprocessing to alleviate numerical ill-conditioning and mitigate noise effects. Case studies on a modified IEEE 39-bus power system show that ambient measurements satisfy the required informativity conditions and that both formulations yield estimates close to model-based references under the tested noise levels. The reusable local ODP estimates also enable modular stability reassessment after controller updates.
\end{abstract}


\section{Introduction}
\label{sec:introduction}

Small-signal stability is a fundamental requirement for secure power system operation. Conventional analysis methods mainly rely on model-based techniques, such as time-domain simulation \cite{stott1979power}, eigenvalue analysis of the system state matrix \cite{sastry1980hierarchical}, and frequency-domain analysis based on the Nyquist criterion \cite{11086493}. However, with the large-scale integration of renewable energy and
distributed energy resources, modern power systems are evolving from
systems dominated by a few large synchronous machines to interconnected
networks composed of numerous heterogeneous devices
\cite{hatziargyriou2020stability}. In such power systems, model-based small-signal stability analysis requires a description of each device's dynamics around a given
operating point and of the interactions among devices through the network \cite{yang2019toward}. In practice, commercial inverters vary in their control structures and
parameter settings, and modeling information such as internal topologies
and parameters can be difficult to obtain from manufacturers for some
devices \cite{men2023artificial}. In addition, maintaining device models over time requires continually checking control configurations and updating the corresponding models. This task becomes more demanding as the number and diversity of
grid-connected devices increase \cite{huang2024robust}. Therefore, there is a need for data-driven small-signal stability assessment methods that reduce reliance on
explicit device models.

Accordingly, data-driven stability analysis of power systems has attracted growing attention. Existing studies in this area mainly follow four approaches: model identification, measurement-based impedance assessment, modal analysis, and
machine-learning-based prediction. Identification-based methods recover dynamic models or estimate system
state matrices \cite{chakrabortty2014allocation,men2025blackbox,huang2026blackbox}, retaining conventional analysis tools but requiring explicit model identification before stability assessment. Measurement-based impedance methods characterize devices from terminal measurements without detailed internal modeling, but stability tests
based on system-wide impedance models require joint frequency-domain analysis of device--network interactions and can become computationally demanding as the system size grows \cite{li2024impedance,11119160}. Modal analysis extracts modal frequencies and damping ratios from ambient and event-driven PMU data to support event identification and small-signal stability assessment \cite{taghipourbazargani2022machine,10123100,yang2026ambient}. However, it primarily characterizes system-level modal properties and does not directly yield reusable device-level descriptions for network-level composition. Machine learning maps operating conditions or measurements to stability margins or labels for rapid assessment, but its reliability depends on training-data coverage and may deteriorate after topology or device changes \cite{10962549,10086543}. Overall, these approaches provide useful model, impedance, modal,
or predictive information, but scalable stability reassessment after
local device changes remains challenging, as system-level analysis
may still need to be repeated.
This motivates a compositional data-driven approach for modular reassessment
without explicit device modeling.

At a given operating point, the network coupling in a power system
can be determined from the network parameters and the steady-state
power-flow solution \cite{kazma2025revisiting}. This provides the interconnection information needed to combine local device analysis results for system-level
stability verification. In power systems, compositional analysis exploits the device-network structure to
decompose system-level stability verification into checks of local
device properties and network interconnection conditions\cite{7934339,sadabadi2017plugandplay}.
This approach allows devices to be analyzed individually and supports
reassessment after local changes by updating the descriptions of
affected devices while reusing the remaining local results.
This reduces reliance on a complete system dynamic model and the need
for repeated system-wide analysis, providing a scalable assessment
approach for large-scale power systems.
Within this approach, dissipativity-based methods
\cite{ojo2023distributed,9667191,10885782}
characterize local device dynamics through dissipation inequalities
and combine these local properties with network coupling conditions. In particular, output differential passivity (ODP) \cite{yang2019distributed} provides local passivity descriptions for compositional system-level assessment and decentralized assessment through local comparison with a network threshold. However, these local dissipativity descriptions usually rely
on analytical device models, making it difficult to obtain
and update them when such models are undisclosed or incomplete.
Whether reusable local ODP matrices can be estimated directly
from local measurement trajectories is therefore central to
extending these methods to data-driven stability assessment
of power systems.

Behavioral systems theory provides a foundation for this purpose.
Willems' fundamental lemma \cite{willems2005note}
shows that finite-length trajectories of a linear system can be
represented by a single measured trajectory satisfying a persistent
excitation condition.
Based on this result, existing studies have further derived conditions
for verifying dissipativity directly from input-state or input-output
trajectories without explicit system models
\cite{koch2020verifying,9551767}.
Our previous work \cite{zhang2026data}
established data-driven ODP certification from input-state trajectories
for power system stability assessment.
However, internal device states may not be measurable in practice.
In addition, applying these methods to ambient measurements in power
systems still poses challenges for computational efficiency and
numerical reliability.
Directly constructed trajectory-based linear matrix inequalities
(LMIs) grow in dimension with the sample size, increasing the
computational burden.
Weak excitation and poor scaling across measurement channels can
lead to numerical ill-conditioning \cite{faleschini2025data},
while measurement noise further affects estimation accuracy.

To address these challenges, this paper extends local ODP
certification for power systems to input-output measurements under ambient
conditions and introduces an equivalent LMI reformulation
and numerical preprocessing to improve computational
efficiency and numerical reliability. The estimated local ODP matrices are used in a compositional stability test, while their scalar indices enable decentralized comparisons with a network threshold. The main contributions of this paper are summarized as follows.

\begin{itemize}
\item We derive a discrete-time ODP condition for LTI systems
to support data-driven assessment from sampled measurements,
and analyze the discretization error margin relative to the
continuous-time ODP condition. Taking this margin into account,
we establish compositional and decentralized small-signal stability
conditions for interconnected power systems.

\item Data-driven LMIs are derived from behaviorally
informative input-state or input-output trajectories under ambient
conditions to estimate local ODP matrices and scalar indices.
To address the computational and numerical challenges,
right-inverse whitening (RIW) reduces the LMI dimension and improves
computational efficiency, while instrumental-variable (IV)
preprocessing alleviates numerical ill-conditioning and mitigates
noise effects.
\end{itemize}

The remainder of this paper is organized as follows.
Section~\ref{sec:preliminaries} introduces the device--network
interconnection and the compositional and decentralized stability
conditions.
Section~\ref{sec:lti-formulations} formulates the continuous-
and discrete-time ODP conditions, analyzes the discretization
error margins, and establishes the corresponding stability conditions.
Section~\ref{sec:input-state} presents the local estimation methods
based on input-state and input-output trajectories.
Section~\ref{sec:unified-behavioral} introduces the unified behavioral representation and right-inverse whitening.
Section~\ref{sec:iv-preprocessing} presents IV preprocessing.
Section~\ref{sec:case-study} evaluates the proposed method
on the IEEE 39-bus system.
Section~\ref{sec:conclusion} concludes this paper.

\section{PROBLEM FORMULATION}
\label{sec:preliminaries}

This section introduces the local device dynamics and their network
interconnection, recalls compositional and decentralized stability
conditions, and formulates the small-signal stability assessment
problem studied in this paper.

\subsection{Device--Network Interconnection Model}
\label{sec:divice-network}
Following \cite{yang2019distributed}, we consider a power system
composed of $N_b$ bus dynamical subsystems interconnected through
the power network. The bus set and transmission-line set are
denoted by $\mathcal V$ and $\mathcal E$, respectively, where
$N_b:=|\mathcal V|$.
At each bus $i\in\mathcal V$, the voltage phasor is denoted by
$V_i\angle\theta_i$, and the active and reactive power injections
are denoted by $P_i$ and $Q_i$, respectively.
Under a fixed angle reference, each device interacts with the network through its power
injections and bus voltage. 
The incremental input and output around a given operating
equilibrium are $u_i:=[-(P_i-P_i^\ast),-(Q_i/V_i-Q_i^\ast/V_i^\ast)]^\top$
and $y_i:=[\theta_i-\theta_i^\ast,V_i-V_i^\ast]^\top$.

Around the same operating equilibrium, the linearized device
dynamics are described by the following continuous-time model:
\begin{equation}
\label{eq:ct-lti-system}
\begin{cases}
\dot{x}_i=A_i x_i+B_i u_i,\\
y_i=C_i x_i,
\end{cases}
\qquad i\in\mathcal V,
\end{equation}
where $x_i=\operatorname{col}(\xi_i-\xi_i^*,\,\theta_i-\theta_i^*,\,V_i-V_i^*)
\in\mathbb R^{n_i}$ collects the device state deviations
from the equilibrium, $u_i,y_i\in\mathbb R^2$,
and $C_i$ is the state-output matrix.
Here, $\xi_i\in\mathbb R^{n_i-2}$ is the auxiliary state variable
which stands for the heterogeneous dynamics of each component.
This model is a linearized version of the general nonlinear
bus dynamics in \cite{yang2019distributed} at the given equilibrium.
It provides a unified input--output representation for heterogeneous
devices with different state dimensions and control structures,
including synchronous generators \cite{stegink2017unifying},
conventional-droop-controlled inverters \cite{6987381}, and
quadratic-droop-controlled inverters \cite{simpson2016voltage},
 with their dynamics captured by the corresponding
system matrices.

The buses are interconnected through the power network characterized
by the admittance matrix $Y_{bus}=G+\mathrm{j}B$.
The corresponding AC power-flow equations are
\begin{equation}
\begin{aligned}
P_i={}&G_{ii}V_i^2+
\sum_{j\in\mathcal{N}_i}V_iV_j
\left(
B_{ij}\sin\theta_{ij}+G_{ij}\cos\theta_{ij}
\right),\\
Q_i={}&-B_{ii}V_i^2-
\sum_{j\in\mathcal{N}_i}V_iV_j
\left(
B_{ij}\cos\theta_{ij}-G_{ij}\sin\theta_{ij}
\right),
\end{aligned}
\label{eq:power_flow}
\end{equation}
where $\mathcal{N}_i$ is the set of buses adjacent to bus $i$,
$\theta_{ij}:=\theta_i-\theta_j$, and $G_{ij}$ and $B_{ij}$ are
the conductance and susceptance entries of the admittance matrix,
respectively.

The power-flow equations and the port definitions determine a static
network mapping $g$, whose Jacobian at the operating equilibrium is
denoted by $J_{\mathrm{net}}:=\left.\frac{\partial g}{\partial y}\right|_{y=0}$.
With the local incremental outputs and inputs stacked as
$y:=\operatorname{col}(y_1,\ldots,y_{N_b})$ and
$u:=\operatorname{col}(u_1,\ldots,u_{N_b})$, respectively,
first-order linearization of the network mapping yields
$u=J_{net}y$.

Interconnecting the local device models with the network yields the small-signal model of the power system

\begin{equation}
\left\{
\begin{aligned}
\dot{x} &= A_Dx+B_Du,\\
y &= C_Dx,\\
u &= J_{net}y.
\end{aligned}
\right.
\label{eq:interconnected_system}
\end{equation}
Here, $x:=\operatorname{col}(x_1,\ldots,x_{N_b})$, and the matrices
$A_D$, $B_D$, and $C_D$ are assembled block-diagonally from the
corresponding local matrices.
The stability problem considered concerns the
equilibrium of this small-signal interconnected system.

\subsection{Compositional and Decentralized Stability Conditions}

\begin{definition}[Generalized ODP \cite{ru2026toward}]
For a given symmetric matrix $\Sigma_i$, if the $i$th local
device admits a continuously differentiable storage function
$S_i(x_i)$ that has a strict local minimum at the equilibrium
$x_i=0$ and satisfies
\begin{equation}
\dot S_i
\leq
u_i^\top\dot y_i-y_i^\top\Sigma_i\dot y_i
\label{eq:matrix_odp}
\end{equation}
along its trajectories, then the device is said to satisfy
matrix-form output differential passivity, denoted by
$\mathrm{ODP}(\Sigma_i)$.
\end{definition}

To state the stability conditions for the interconnected power
system, collect the local ODP matrices into the block-diagonal
matrix
\begin{equation}
D_\Sigma
:=
\operatorname{blkdiag}
\left(\Sigma_1,\ldots,\Sigma_{N_b}\right).
\label{eq:aggregated_odp_matrix}
\end{equation}

The Lyapunov function for the interconnected power system combines the local
storage functions of the devices with a network energy term. In the
stability analysis of lossy power systems, numerical energy function
methods have been used to approximate the path-dependent integrals
arising from transfer conductances~\cite{chiang1995direct,guo2025efficient}.

Following the numerical energy function approximation
in~\cite{yang2019distributed}, we construct a quadratic network
energy term around the operating equilibrium:
\[
W_N(y):=\frac{1}{2}y^\top H_{net}y,
\qquad
H_{net}
:=-\frac{1}{2}\bigl(J_{net}+J_{net}^\top\bigr).
\]
Here, $H_{net}$ is the Hessian of the approximate network
energy. Combining this energy term with the local device storage
functions yields the following stability condition.

\begin{lemma}[Compositional small-signal stability condition]
\label{lem:compositional-stability}
Consider the small-signal interconnected power system~\eqref{eq:interconnected_system}
under the above numerical energy function approximation.
If every local device satisfies ODP($\Sigma_i$) and
\begin{equation}
H_{net}+D_\Sigma\succ0,
\label{eq:compositional_condition}
\end{equation}
then the equilibrium of the interconnected power system is
Lyapunov stable.
\end{lemma}

\textit{Proof: See Appendix~\ref{app:proof-compositional}.}

\begin{corollary}[Decentralized small-signal stability condition]
\label{cor:decentralized-stability}
Under the remaining assumptions of
Lemma~\ref{lem:compositional-stability},
let $s_i$ be a scalar index satisfying
$\Sigma_i\succeq s_i I_{m_i}$, where $m_i=\dim(y_i)$, and define
$\sigma_{net}:=-\lambda_{\min}(H_{net})$.
If
\begin{equation}
s_i>\sigma_{net},
\qquad \forall i\in\mathcal{V},
\label{eq:distributed_stability_condition}
\end{equation}
then the equilibrium is Lyapunov stable.
\end{corollary}

\textit{Proof: See Appendix~\ref{app:proof-decentralized}.}

The compositional condition retains the spatial and directional
information of the ODP matrices and is generally less conservative,
whereas the decentralized condition requires only local scalar
comparisons with the network threshold. Lemma~\ref{lem:compositional-stability} and
Corollary~\ref{cor:decentralized-stability} provide the
Lyapunov-stability guarantee underlying the proposed data-driven
assessment framework.

For a given network and operating point, $H_{net}$ and
$\sigma_{net}$ can be computed from the network parameters and
equilibrium power-flow solution \cite{kazma2025revisiting}.
In practice, the grid operator typically has access to this network
information, but detailed control models of some connected devices
may be unavailable. Therefore, this paper estimates local ODP matrices
and scalar indices from local measurement trajectories and combines
these estimates with the network coupling information to assess the
small-signal stability of the interconnected power system.

\subsection{Assessment Problem}

The stability conditions in \eqref{eq:compositional_condition}
and \eqref{eq:distributed_stability_condition} are formulated
in continuous time, whereas measurements are available as sampled
sequences. We thus need a corresponding discrete-time ODP condition, together
with an analysis of the discretization error margin relative to its
continuous-time counterpart.

Building on this relationship, this paper investigates how to combine
local discrete-time ODP descriptions with network coupling information
to establish compositional and decentralized small-signal stability
conditions and how to estimate the required ODP matrices and scalar
indices from local measurement trajectories without explicit device
models. Input-state and input-output measurement settings are
considered, depending on whether internal states are available.
The computational and numerical challenges arising in the
resulting data-driven conditions are also addressed.

\section{ODP Conditions and Stability Analysis}
\label{sec:lti-formulations}

This section presents the continuous-time ODP condition in LMI form
and introduces a discrete-time ODP definition together with its LMI
characterization for sampled linear time-invariant (LTI) systems. The discretization error margin
relative to the continuous-time condition is then analyzed.
Finally, accounting for this margin, a stability theorem is established
that combines local discrete-time ODP descriptions with network
coupling information to provide compositional and decentralized
small-signal stability conditions for interconnected power systems.

\subsection{ODP for a Continuous-Time LTI System}

For a generic local device, we omit the device subscript $i$
and denote the state and port dimensions by $n$ and $m$,
respectively.
With a quadratic storage function, the ODP condition for the
continuous-time LTI device model \eqref{eq:ct-lti-system}
can be equivalently expressed as the following LMI.

\begin{lemma}
\label{lem:ct-lti}
For a given symmetric matrix $\Sigma$, the continuous-time LTI
system~\eqref{eq:ct-lti-system} admits a quadratic storage function
$S(x)=x^\top Px$ satisfying ODP($\Sigma$) if and only if there exists
a symmetric positive-definite matrix $P\succ0$ satisfying the following
LMI:
\begin{equation}\label{eq:ct-lti}
M_1 - \frac{1}{2}(M_2 + M_2^\top) \preceq 0,
\end{equation}
where the corresponding block matrices are defined as:
\begin{equation}
M_1 := \begin{bmatrix} A^\top P + PA & PB \\ * & 0 \end{bmatrix},
\end{equation}
\begin{equation}
M_2 := \begin{bmatrix} -C^\top \Sigma CA & -C^\top \Sigma CB \\ CA & CB \end{bmatrix}.
\end{equation}
\end{lemma}
\textit{Proof: See Appendix~\ref{app:proof-ct-lmi}.}

\subsection{ODP for a Discrete-Time LTI System}
Data-driven verification frameworks inherently operate on sampled measurement sequences rather than continuous-time trajectories. Accordingly, the continuous-time system~\eqref{eq:ct-lti-system} is discretized, yielding the following discrete-time LTI model:
\begin{equation}\label{eq:dt-lti-system}
\begin{cases}
x_{k+1} = \bar{A}x_k + \bar{B}u_k \\
y_k = Cx_k
\end{cases},
\end{equation}
where $x_k \in \mathbb{R}^n$, $u_k \in \mathbb{R}^m$, and
$y_k \in \mathbb{R}^m$ denote the state, input, and output vectors
at the $k$-th sampling instant, respectively.
The sampling interval is $h>0$.
For a first-order consistent discretization, the discrete-time
matrices satisfy the following relations as $h\to0$:
\begin{equation}
\label{eq:exact-discretization}
\bar A=I+hA+O(h^2),\qquad \bar B=hB+O(h^2).
\end{equation}
Using the sampled output increment $\Delta y_k:=y_{k+1}-y_k$,
the discrete-time ODP condition is defined as follows.

\begin{definition}
For a given symmetric matrix $\Sigma^d$, suppose there exists
a positive-definite storage function $S(x_k)$ for which the following
inequality holds along every trajectory of the dt-LTI
system~\eqref{eq:dt-lti-system}
\begin{equation}
\label{eq:dt-odp-dissipation}
S(x_{k+1})-S(x_k)
\leq
(u_k-\Sigma^d y_k)^\top(y_{k+1}-y_k).
\end{equation}
Then the system is said to satisfy ODP($\Sigma^d$).
\end{definition}

The following theorem gives an LMI characterization of ODP
with a quadratic storage function.

\begin{theorem}\label{thm:dt-odp-lmi}
The dt-LTI system~\eqref{eq:dt-lti-system} satisfies ODP($\Sigma^d$)
with a quadratic storage function $S(x_k)=x_k^\top P x_k$ if and
only if there exists $P\succ0$ such that the following LMI holds:
\begin{equation}\label{eq:dt-odp-lmi}
\bar{M}_{1} - \frac{1}{2}(\bar{M}_{2} + \bar{M}_{2}^\top) \preceq 0,
\end{equation}
where the corresponding block matrices are defined as:
\begin{equation}
\label{eq:M1}
\bar{M}_{1} := \begin{bmatrix} \bar{A}^\top P \bar{A} - P & \bar{A}^\top P \bar{B} \\ * & \bar{B}^\top P \bar{B} \end{bmatrix},
\end{equation}
\begin{equation}
\label{eq:M2}
\bar{M}_{2} := \begin{bmatrix} -C^\top \Sigma^d C(\bar{A} - I) & -C^\top \Sigma^d C\bar{B} \\ C(\bar{A} - I) & C\bar{B} \end{bmatrix}.
\end{equation}
\end{theorem}

\textit{Proof: See Appendix~\ref{app:proof-dt-lmi}.}

Theorem~\ref{thm:dt-odp-lmi} converts the discrete-time ODP condition into a computable LMI, providing a model-based reference for the subsequent data-driven criteria.

\subsection{Discretization Error Margins and Stability Conditions}

Local ODP descriptions are obtained from sampled measurements,
whereas the physical grid evolves in continuous time.
It is therefore necessary to examine the relationship between
the discrete- and continuous-time ODP conditions and account
for the resulting discretization error margins in stability assessment.

\begin{proposition}
\label{prop:discretization-error}
For fixed $P\succ0$ and a symmetric matrix $\Sigma^d$, write
the continuous- and discrete-time ODP conditions in
\eqref{eq:ct-lti} and \eqref{eq:dt-odp-lmi} as
$R_c(P,\Sigma^d)\preceq0$ and
$R_d(P,\Sigma^d;h)\preceq0$, respectively.
If the discrete-time matrices satisfy \eqref{eq:exact-discretization},
then, as $h\to0$,
\[
h^{-1}R_d(P,\Sigma^d;h)
=
R_c(P,\Sigma^d)+E_h(P,\Sigma^d),
\]
where $\|E_h(P,\Sigma^d)\|_2=O(h)$.
\end{proposition}

\textit{Proof: See Appendix~\ref{app:proof-discretization}.}

The above proposition shows that the difference between
$h^{-1}R_d$ and $R_c$ vanishes as $h\to0$. For finite sampling intervals, using local ODP estimates obtained
from sampled data for continuous-time stability assessment still
requires a discretization error margin to avoid overestimating the stability margin in
device--network matching.

We now present the main stability result for the interconnected
power system based on local discrete-time ODP.

To account for the discretization discrepancy, assume that, for each
device $i\in\mathcal V$, the continuous-time local system
\eqref{eq:ct-lti-system} admits an ODP matrix $\Sigma_i$ satisfying
$\Sigma_i\succeq\Sigma_i^d-\varepsilon_i I_{m_i}$, where
$\varepsilon_i\ge0$ is a discretization error margin.
Define $D_{\Sigma^d}:=\operatorname{blkdiag}(\Sigma_i^d)$ and
$D_{\varepsilon}:=\operatorname{blkdiag}(\varepsilon_i I_{m_i})$.

\begin{theorem}\label{thm:main}
Suppose that, for each $i\in\mathcal V$, the local discrete-time
system satisfies ODP($\Sigma_i^d$), with
$\Sigma_i^d\succeq s_i I_{m_i}$.
Under the above discretization error margins:

1) if $H_{net}+D_{\Sigma^d}\succ D_{\varepsilon}$, the
compositional Lyapunov-stability condition holds;

2) if $s_i-\sigma_{net}>\varepsilon_i$ for all $i\in\mathcal V$, the
decentralized Lyapunov-stability condition holds.
\end{theorem}

\textit{Proof: See Appendix~\ref{app:proof-main}.}

Theorem~\ref{thm:main} combines local discrete-time ODP descriptions
with network coupling information to establish continuous-time
small-signal stability conditions under the assumed discretization
error margins for individual devices. In practice, continuous-time
device models may be unavailable, making these margins difficult to
verify for each device. In this case, the estimated stability margins
obtained at different sampling intervals can be compared at the same
operating point, and their variation can be used to set an empirical
decision buffer. A stability test is considered satisfied when its
estimated margin exceeds this buffer.
The stability guarantee of Theorem~\ref{thm:main} still requires
the discretization error margins for individual devices to hold.

\section{Data-Driven Conditions for Local ODP Verification}
\label{sec:input-state}

Building on the discrete-time ODP condition, this section
establishes data-driven LMI criteria for local ODP assessment
from measurement trajectories. The input-state setting is
considered first, followed by an extended-state construction
for assessment using only input-output measurements.

\subsection{Input-State Data and Rank Condition}

Our objective is to verify the discrete-time ODP property without explicitly identifying the system matrices $(\bar{A}, \bar{B})$. We collect an input sequence $\{u_k\}_{k=0}^{N-1}$ and the state sequence $\{x_k\}_{k=0}^{N}$, then construct the following data matrices:
\[
\begin{alignedat}{2}
X
&:= \begin{bmatrix}
x_0 & \cdots & x_{N-1}
\end{bmatrix},
&\qquad
Y
&:= \begin{bmatrix}
y_0 & \cdots & y_{N-1}
\end{bmatrix},\\
X_+
&:= \begin{bmatrix}
x_1 & \cdots & x_N
\end{bmatrix},
&
Y_+
&:= \begin{bmatrix}
y_1 & \cdots & y_N
\end{bmatrix},\\
U
&:= \begin{bmatrix}
u_0 & \cdots & u_{N-1}
\end{bmatrix}.
&&
\end{alignedat}
\]

The proposed data-driven framework relies on measured input-state trajectories~\cite{koch2020verifying}. A fundamental prerequisite for this approach is that the collected data must be sufficiently informative to characterize the underlying system dynamics. This requirement is formalized by the following full-row-rank condition~\cite{willems2005note}:
\begin{equation}\label{eq:is-rank-condition}
\operatorname{rank} \begin{bmatrix} X \\ U \end{bmatrix} = n + m.
\end{equation}

This rank condition can be guaranteed if the input sequence is persistently exciting of a sufficiently high order. Given the finite input sequence $\{u_k\}_{k=0}^{N-1}$, we construct the corresponding Hankel matrix of depth $L$ as follows:
\begin{equation}
H_L(u) := \begin{bmatrix} 
u_0 & u_1 & \cdots & u_{N-L} \\ 
u_1 & u_2 & \cdots & u_{N-L+1} \\ 
\vdots & \vdots & \ddots & \vdots \\ 
u_{L-1} & u_L & \cdots & u_{N-1} 
\end{bmatrix}.
\end{equation}

\begin{definition}[Persistent Excitation ~\cite{koch2020verifying}]
An input sequence is persistently exciting of order $L$ when its
depth-$L$ Hankel matrix has full row rank, equivalently,
$\operatorname{rank}(H_L(u))=mL$.
\end{definition}

If the pair $(\bar{A},\bar{B})$ is controllable, the fundamental lemma implies that the rank condition in~\eqref{eq:is-rank-condition} is guaranteed when the input sequence is persistently exciting of order $n+1$~\cite{9551767}.

\subsection{Input-State Data-Driven ODP Criterion}

Using discrete-time input-state trajectories avoids explicit identification of the system matrices. Provided the data satisfy the informativity rank condition, the following theorem establishes a data-driven LMI criterion for ODP verification.

\begin{theorem}\label{thm:is-data-odp}
Let $\{u_k\}_{k=0}^{N-1}$ and $\{x_k\}_{k=0}^N$ be input-state
trajectories of the dt-LTI system~\eqref{eq:dt-lti-system} satisfying
\eqref{eq:is-rank-condition}. For a symmetric matrix $\Sigma^d$, the
system satisfies ODP($\Sigma^d$) with a quadratic storage function
$S(x_k)=x_k^\top P x_k$ if and only if there exists
$P=P^\top\succ0$ such that
\begin{equation}\label{eq:is-data-lmi}
X_+^\top P X_+ - X^\top P X - \frac{1}{2}(M_3 + M_3^\top) \preceq 0,
\end{equation}
where $M_3:=(U-\Sigma^d CX)^\top(CX_+-CX)$.
\end{theorem}

\textit{Proof: See Appendix~\ref{app:proof-is-data}.}

Theorem~\ref{thm:is-data-odp} enables direct local ODP certification
when states are measurable without identifying $\bar{A}$ and $\bar{B}$; if $C$
is unknown but $Y$ and $Y_+$ are available, they replace $CX$ and
$CX_+$, respectively. 

\subsection{Extended State from Input-Output Data}
\label{sec:input-output}

When internal states are unavailable, consider an input-output
trajectory $\{u_k\}_{k=0}^{N-1}$, $\{y_k\}_{k=0}^{N}$ of
\eqref{eq:dt-lti-system}.
The system is assumed to be a minimal realization, and its
order $n$ is known from structural device information.
To derive a data-driven ODP criterion, an extended realization
with the same reachable input-output behavior is constructed
from past inputs and outputs~\cite{9551767}.

\begin{definition}
The minimum lag $\underline l$ of
\eqref{eq:dt-lti-system} is the smallest $l\in\mathbb N_{+}$ for which the observability matrix $\mathcal O_l:=\left[C^\top\ (C\bar{A})^\top\ \cdots\ (C\bar{A}^{l-1})^\top\right]^\top$ has rank $n$.
\end{definition}

For any $l\geq\underline l$, define the extended state
\begin{equation}
\xi_k:=
\left[
u_{k-l}^\top\ \cdots\ u_{k-1}^\top\
y_{k-l}^\top\ \cdots\ y_{k-1}^\top
\right]^\top .
\label{eq:io_extended_state}
\end{equation}

\begin{lemma}\label{lag}
Let $l\geq\underline l$. Then there exist matrices
$\bar{\widetilde A},\bar{\widetilde B},\widetilde C$ such that, for
$k=l,\ldots,N-1$, the measured input-output trajectory admits the
extended realization
\begin{equation}
\xi_{k+1}=\bar{\widetilde A}\xi_k+\bar{\widetilde B}u_k,
\qquad
y_k=\widetilde C\xi_k.
\label{eq:extended-lti}
\end{equation}
Conversely, every input-output trajectory of
\eqref{eq:extended-lti} initialized from a reachable extended state is
also an input-output trajectory of \eqref{eq:dt-lti-system}.
\end{lemma}

\begin{proof}
This follows from \cite[Lemma~2]{9551767}, noting that the output of
\eqref{eq:dt-lti-system} is independent of the current input.
\end{proof}

Let $\mathcal X_\xi^r$ denote the reachable subspace of the extended
realization, namely the set of extended states generated by valid
input-output histories. It need not be computed explicitly because,
under the informativity condition ~\eqref{eq:io_behavioral_rank}, the columns of $\Xi_l$ span this subspace. By Lemma~\ref{lag}, the original and extended systems have the same reachable input-output behavior.

For each candidate lag $l$, collect
\begin{alignat*}{2}
&\Xi_l:=[\xi_l\ \xi_{l+1}\,\cdots\,\xi_{N-1}],\quad
&&\Xi_{l,+}:=[\xi_{l+1}\ \xi_{l+2}\,\cdots\,\xi_N],\\
&U_{\Xi,l}:=[u_l\ u_{l+1}\,\cdots\,u_{N-1}],\quad
&&Y_{\Xi,l}:=[y_l\ y_{l+1}\,\cdots\,y_{N-1}],\\
&Y_{\Xi,l,+}:=[y_{l+1}\ y_{l+2}\,\cdots\,y_N].
\end{alignat*}
The extended realization may be nonminimal, so
$[\Xi_l^\top\ U_{\Xi,l}^\top]^\top$ need not have full row rank. It
only needs to span all reachable state-input pairs of the extended
realization \cite{9551767}.

\subsection{Behavioral Informativity and Input-Output ODP Criterion}

Although $l=n$ is sufficient, it may produce an unnecessarily large extended state. If the input is persistently exciting of order $n+l+1$, the Fundamental Lemma gives the behavioral dimension relation~\cite{9551767}
\begin{equation}
\operatorname{rank}
\begin{bmatrix}
\Xi_l\\ U_{\Xi,l}
\end{bmatrix}
=
(l+1)m+\operatorname{rank}(\mathcal O_l).
\label{eq:io_behavioral_rank}
\end{equation}
Combining \eqref{eq:io_behavioral_rank} with the definition of $\underline l$, the minimum valid lag is selected as
\begin{equation}
l_d:=
\min\left\{
l\in\mathbb N_+:
\operatorname{rank}
\begin{bmatrix}
\Xi_l\\U_{\Xi,l}
\end{bmatrix}
=n+(l+1)m
\right\}.
\label{eq:data_driven_lag_selection}
\end{equation}

Here, $n$ is the known device order obtained from device-type or controller-structure information and is used only to set the rank target. Hereafter, we set $l=l_d$ and omit the subscript $l$ from the data matrices.

With $l=l_d$ selected by~\eqref{eq:data_driven_lag_selection}, the extended-state data yield the following input-output ODP criterion.

\begin{theorem}\label{thm:io-data-odp}
Consider an input-output trajectory $\{u_k\}_{k=0}^{N-1}$, $\{y_k\}_{k=0}^{N}$ of the minimal dt-LTI system \eqref{eq:dt-lti-system}. Let $l=l_d$ be selected by \eqref{eq:data_driven_lag_selection}. For a symmetric matrix
$\Sigma^d$, if there exists $P=P^\top\succ0$ such that
\begin{equation}
\Xi_+^\top P\Xi_+
-\Xi^\top P\Xi
-\frac12(M_4+M_4^\top)
\preceq0,
\label{eq:io-odp-lmi}
\end{equation}
where $M_4:=(U_\Xi-\Sigma^dY_\Xi)^\top
(Y_{\Xi,+}-Y_\Xi)$, then the system in \eqref{eq:dt-lti-system} satisfies
ODP($\Sigma^d$) with a quadratic storage function
$S_x(x_k)=x_k^\top P_x x_k$.
\end{theorem}

\textit{Proof: See Appendix~\ref{app:proof-io-data}.}

Theorem~\ref{thm:io-data-odp} establishes direct local ODP certification from terminal input-output trajectories, without internal-state measurements or system matrices.

\section{Unified Behavioral Representation and Efficient Computation}
\label{sec:unified-behavioral}

The input-state and input-output formulations share the same
algebraic structure, enabling a unified behavioral representation.
However, the resulting LMIs~\eqref{eq:is-data-lmi} and ~\eqref{eq:io-odp-lmi} grow in dimension with the number of
samples, while data redundancy and scale differences can cause
numerical ill-conditioning.
We introduce a right-inverse whitening (RIW) technique that
preserves equivalence while improving conditioning and reducing
dimensionality.
Finally, local ODP matrix estimates are obtained via semidefinite
programming (SDP).

\subsection{Unified Behavioral Data Representation}
\label{subsec:unified-behavioral-representation}

The input-state criterion in \eqref{eq:is-data-lmi} and the input-output criterion in \eqref{eq:io-odp-lmi} have the same algebraic structure after the data columns are aligned. To write them in a unified form, let $\eta_k$ denote the state used in the data-driven LMI. For the input-state case, $\eta_k=x_k$. For the input-output case, we re-index the aligned samples
$(\xi_{k+l},u_{k+l},y_{k+l})$ as $(\eta_k,u_k,y_k)$.

With this notation, collect
\begin{align*}
    & \mathcal{X}:=[\eta_0 \ \eta_1 \ \cdots \ \eta_{N_d-1}], \qquad
    && \mathcal{X}_+:=[\eta_1 \ \eta_2 \ \cdots \ \eta_{N_d}],\\
    & \mathcal{U}:=[u_0 \ u_1 \ \cdots \ u_{N_d-1}], \qquad
    && \mathcal{Y}:=[y_0 \ y_1 \ \cdots \ y_{N_d-1}],\\
    & \mathcal{Y}_+:=[y_1 \ y_2 \ \cdots \ y_{N_d}].
\end{align*}
Here, $(\mathcal{X},\mathcal{X}_+,\mathcal{U},\mathcal{Y},\mathcal{Y}_+)
=(\Xi,\Xi_+,U_\Xi,Y_\Xi,Y_{\Xi,+})$ for the input-output formulation, where $N_d$ denotes the number of aligned data transitions.
The output-related data matrices are constructed as
$\mathcal{G}=C\mathcal{X}$ and $\mathcal{G}_+=C\mathcal{X}_+$
for the input-state case, while
$\mathcal{G}=\mathcal{Y}$ and $\mathcal{G}_+=\mathcal{Y}_+$
for the input-output case.
For $Z:=[\mathcal{X}^\top\ \mathcal{U}^\top]^\top$,
both \eqref{eq:is-data-lmi} and \eqref{eq:io-odp-lmi} can be written as
\begin{equation}
  \begin{aligned}
    \mathcal F(P,\Sigma^d)
    :={}&
    \mathcal X_+^\top P\mathcal X_+
    -\mathcal X^\top P\mathcal X \\
    &-\frac{1}{2}\bigl(\mathcal M(\Sigma^d)+\mathcal M(\Sigma^d)^\top\bigr)
    \preceq0,
  \end{aligned}
  \label{eq:unified_data_lmi}
\end{equation}
where $\mathcal M(\Sigma^d):=
(\mathcal U-\Sigma^d\mathcal G)^\top
(\mathcal G_+-\mathcal G)$.

The rank requirement on $Z$ depends on the formulation. In the input-state case, it reduces to \eqref{eq:is-rank-condition}. In the input-output case, the extended state may be nonminimal, and the required rank is determined by the behavioral rank condition in \eqref{eq:data_driven_lag_selection}, rather than the full row dimension of $Z$.

\subsection{Right-Inverse Whitening Reformulation}
\label{subsec:right_inverse_whitening}

Although the unified LMI~\eqref{eq:unified_data_lmi} is model-free,
solving it directly on all columns of $Z$ can be computationally
expensive and numerically ill-conditioned.
For a trajectory containing $N_d$ data transitions, the LMI has
dimension $N_d$.
Let $r:=\operatorname{rank}(Z)$ denote the dimension of the behavioral
space spanned by the columns of $Z$, i.e., the number of independent
behavioral directions represented by the data.
Typically, $r$ is much smaller than $N_d$.
Once the data span the required behavioral space, additional columns
increase the LMI dimension without introducing new independent
behavioral directions.
Moreover, weak excitation and poor scaling across measurement channels
can lead to widely differing nonzero singular values of $Z$,
further impairing numerical conditioning.
We therefore use RIW to reformulate the LMI on a compact orthonormal
basis of this behavioral space, retaining all independent directions
while removing redundant coordinates and normalizing their scales.
This reduces the LMI dimension to $r$ and improves computational
efficiency and numerical conditioning.

\begin{definition}[RIW matrix]
Let $Z=Q_rS_rV_r^\top$ be the compact singular value
decomposition of $Z$, where $r=\operatorname{rank}(Z)$
and $S_r\succ0$. The RIW matrix associated with $Z$
is defined as
\begin{equation}
W := V_rS_r^{-1}.
\label{eq:riw-matrix}
\end{equation}
\end{definition}

By construction,
\begin{equation}
    ZW=Q_r,\qquad
    W^\top Z^\top ZW=I_r .
    \label{eq:whitening_identity}
\end{equation}
Thus, multiplying the data matrices by $W$ replaces the original data columns with an orthonormal basis of the behavioral space. Since $(Q_r^\top Z)W=I_r$, the matrix $W$ is a right inverse of the reduced behavioral matrix $Q_r^\top Z$ \cite{de2019formulas}.

Applying the same transformation to all data matrices gives
\begin{align*}
\widehat{\mathcal X}&:= \mathcal X W,&
\widehat{\mathcal X}_+ &:= \mathcal X_+ W,\\
\widehat{\mathcal U}   &:= \mathcal U W,   &
\widehat{\mathcal G}   &:= \mathcal G W,   &
\widehat{\mathcal G}_+ &:= \mathcal G_+ W .
\end{align*}
The whitened data-driven ODP LMI is
\begin{equation}
\begin{aligned}
\widehat{\mathcal F}(P,\Sigma^d):={}&
\widehat{\mathcal X}_+^\top P\widehat{\mathcal X}_+
-\widehat{\mathcal X}^\top P\widehat{\mathcal X}  \\
&-\frac{1}{2}
\left(\widehat{\mathcal M}(\Sigma^d)
+\widehat{\mathcal M}(\Sigma^d)^\top\right)
\preceq0,
\end{aligned}
\label{eq:whitened_lmi}
\end{equation}
where
\begin{equation}
    \widehat{\mathcal M}(\Sigma^d)
    :=
    (\widehat{\mathcal U}-\Sigma^d\widehat{\mathcal G})^\top
    (\widehat{\mathcal G}_+-\widehat{\mathcal G}) .
    \label{eq:whitened_M}
\end{equation}

The following result shows that the above reformulation does not change the ODP condition when all informative singular directions are retained.

\begin{theorem}[Lossless right-inverse whitening]\label{thm:riw}
Assume that the measured data are generated by the corresponding LTI system and that $Z$ spans the required behavioral space with rank $r$. Then the original LMI~\eqref{eq:unified_data_lmi} and the whitened LMI~\eqref{eq:whitened_lmi} are equivalent, i.e.,
\begin{equation}
    \mathcal F(P,\Sigma^d)\preceq0
    \quad\Longleftrightarrow\quad
    \widehat{\mathcal F}(P,\Sigma^d)\preceq0 .
    \label{eq:whitening_equivalence}
\end{equation}
\end{theorem}

\textit{Proof: See Appendix~\ref{app:proof-riw}.}

In the input-state case, the rank condition makes $Z$ full row rank, so RIW only changes coordinates. In the input-output case, the extended state may be nonminimal, so RIW removes redundant behavioral coordinates without changing the ODP condition.

\subsection{Matrix ODP Estimation}

Using the LMI~\eqref{eq:whitened_lmi} obtained after RIW,
we jointly estimate the local ODP matrix $\Sigma^d$ and
scalar index $s$ through the following semidefinite program:

\begin{equation}
\begin{aligned}
\underset{\Sigma^d,\,P,\,s}{\max}\quad & s\\
\mathrm{s.t.}\quad
& \widehat{\mathcal F}(P,\Sigma^d)\preceq0,\\
& P=P^\top\succ0,\\
& \Sigma^d=(\Sigma^d)^\top,\\
& \Sigma^d\succeq sI_m.
\end{aligned}
\label{eq:whitened_sdp}
\end{equation}

These local estimates are used in Theorem~\ref{thm:main}. For device $i$, \eqref{eq:whitened_sdp} yields the matrix
estimate $\Sigma_i^d$ and scalar index $s_i$.
The full matrix retains directional information for
compositional assessment, while the scalar index supports
decentralized comparison with the network threshold.

\section{Mitigating Noise Effects and Improving Numerical Conditioning}
\label{sec:iv-preprocessing}

The preceding data-driven ODP criteria~\eqref{eq:is-data-lmi} and ~\eqref{eq:io-odp-lmi} assume noise-free data,
but weak excitation and poor data scaling can still cause
numerical ill-conditioning. Measurement noise can further affect
behavioral subspace extraction and ODP estimation.
We therefore apply instrumental-variable (IV) preprocessing
jointly to the complete LMI data block before RIW to improve
numerical conditioning and mitigate noise effects.

\subsection{Instrument Selection and Data Alignment}
In the unified representation of
Section~\ref{subsec:unified-behavioral-representation},
the noisy data corresponding to $\eta_k,u_k,\gamma_k$
are denoted by
$\widetilde\eta_k,\widetilde u_k,\widetilde\gamma_k$,
respectively. Here, $\gamma_k=Cx_k$ for the input-state case and
$\gamma_k=y_k$ for the input-output case.
To construct the instruments and jointly preprocess the
data required for ODP estimation, we assemble the
behavioral column and the complete LMI data column
as follows:

\[
\begin{aligned}
\widetilde z_k
&:=\operatorname{col}(\widetilde\eta_k,\widetilde u_k),\\
\widetilde d_k
&:=\operatorname{col}
(\widetilde\eta_k,\widetilde\eta_{k+1},
\widetilde u_k,\widetilde\gamma_k,\widetilde\gamma_{k+1}).
\end{aligned}
\]

We use lagged behavioral measurements as
IVs~\cite{soderstrom2002instrumental,van2022data}.
The lag $\tau$ is selected so that the instruments are
weakly correlated with the measurement noise in the data
being processed while remaining sufficiently correlated
with the underlying noise-free behavior.
These requirements correspond to IV exogeneity and
relevance, respectively.
For the input-output case, the lag should also avoid
sharing raw measurement samples between the instruments
and the extended state being processed.
The available sample size must also be considered when
selecting $\tau$.

We then pair each data column $\widetilde d_k$ with
the earlier behavioral column $\widetilde z_{k-\tau}$.
The aligned data form
$\widetilde D_a:=
[\widetilde d_\tau\ \cdots\ \widetilde d_{N_d-1}]$
and
$\widetilde Z_{\mathrm{iv}}:=
[\widetilde z_0\ \cdots\ \widetilde z_{N_d-\tau-1}]$.
The matching data columns are collected in
$\widetilde Z_a:=
[\widetilde z_\tau\ \cdots\ \widetilde z_{N_d-1}]$.

All three matrices contain $N_{\mathrm{iv}}=N_d-\tau$
columns.
We use $\widetilde Z_{\mathrm{iv}}$ to construct the joint
preprocessing transformation applied to $\widetilde D_a$,
and $\widetilde Z_a$ to check the numerical rank required
by RIW after preprocessing.

\subsection{IV Preprocessing and Assessment Procedure}

To construct the IV filter, we compute the compact singular
value decomposition
$\widetilde Z_{\mathrm{iv}}
=Q_{\mathrm{iv}}S_{\mathrm{iv}}V_{\mathrm{iv}}^\top$.
The columns of $V_{\mathrm{iv}}$ form an orthonormal basis
for the row space of the instrument matrix.
We use the corresponding singular values to set the
attenuation weights in the following Tikhonov filter:
\begin{equation}
\Pi_{\mathrm{iv},\lambda}
:=
V_{\mathrm{iv}}
\operatorname{diag}\!\left(
\frac{\rho_j^2}{\rho_j^2+\lambda}
\right)
V_{\mathrm{iv}}^\top,
\qquad \lambda>0,
\label{eq:regularized_ivop}
\end{equation}
where $\rho_j$ is the corresponding singular value and
$\lambda$ controls the regularization strength.
Components associated with smaller singular values are
attenuated more strongly.
For $\lambda>0$, this matrix is a symmetric contraction;
as $\lambda\to0$, it approaches the orthogonal projector
onto the row space of the instrument matrix.

\begin{figure}[!t]
    \centering
    \includegraphics[width=0.8\columnwidth]{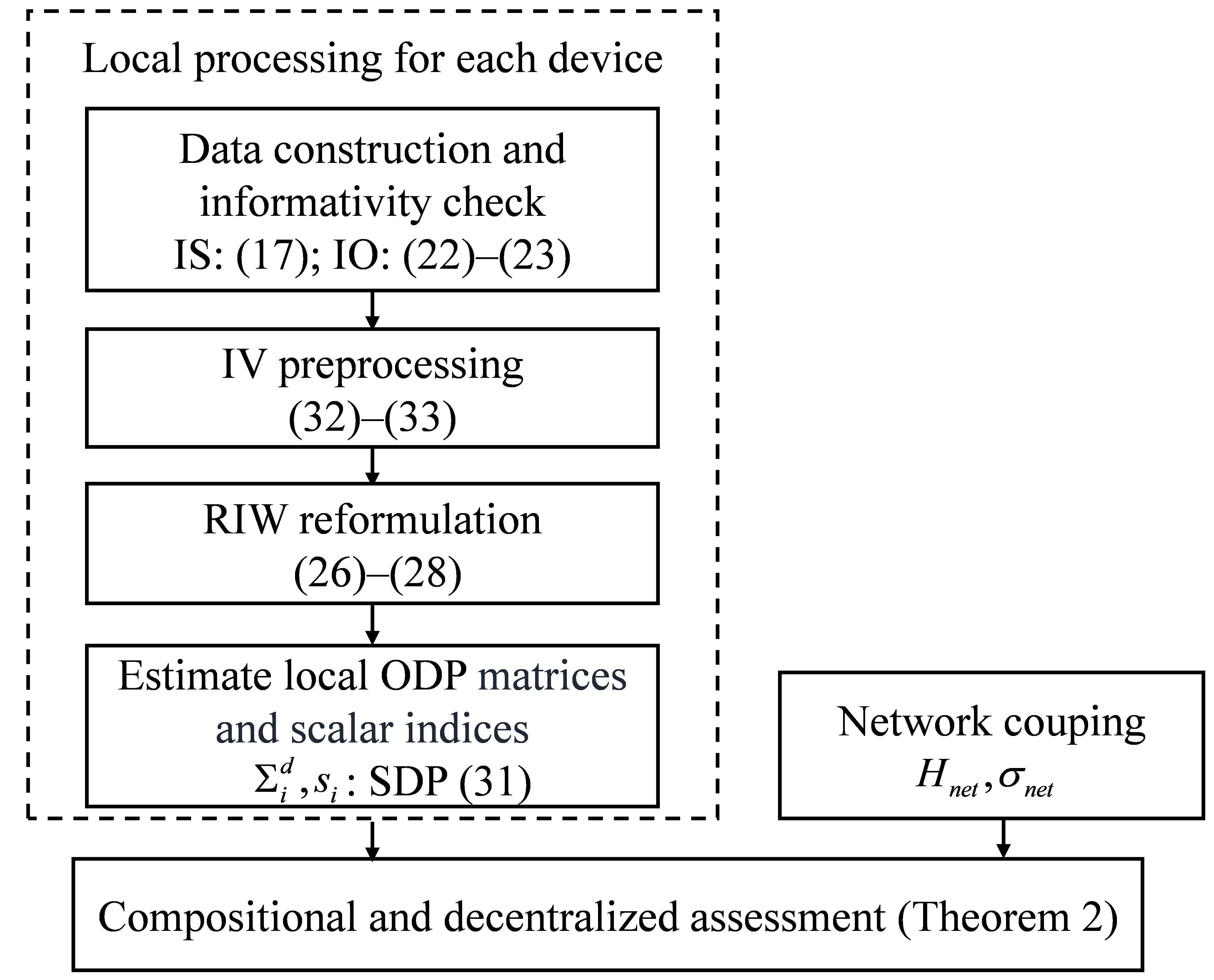}
    \caption{Overall procedure of the proposed assessment method.}
    \label{fig:assessment_flowchart1}
\end{figure}

Full filtering may also suppress weak but informative
components. We therefore introduce a blending parameter
$\alpha$ to adjust the preprocessing strength:
\begin{equation}
\begin{aligned}
\Pi_{\alpha,\lambda}
&:=(1-\alpha)I+\alpha\Pi_{\mathrm{iv},\lambda},\\
D_{\mathrm{iv}}
&:=\widetilde D_a\Pi_{\alpha,\lambda},
\end{aligned}
\label{eq:blended_ivop}
\end{equation}
where $\alpha\in[0,1]$.
We repartition the preprocessed data as
\[
D_{\mathrm{iv}}
=\operatorname{col}\!\left(
\mathcal X_{\mathrm{iv}},
\mathcal X_{\mathrm{iv},+},
\mathcal U_{\mathrm{iv}},
\mathcal G_{\mathrm{iv}},
\mathcal G_{\mathrm{iv},+}
\right).
\]
These blocks replace
$(\mathcal X,\mathcal X_+,\mathcal U,\mathcal G,\mathcal G_+)$,
respectively, in the unified LMI~\eqref{eq:unified_data_lmi}.
After checking the required rank, we apply RIW and solve
SDP~\eqref{eq:whitened_sdp} to obtain local ODP matrices
and scalar indices.
These estimates are combined with network coupling
information for compositional and decentralized stability
assessment, respectively. The overall procedure is summarized in
Fig.~\ref{fig:assessment_flowchart1}.

In~\eqref{eq:blended_ivop}, $\alpha=0$ retains the aligned
original data; $\alpha=1$ applies the full regularized filter, while
$0<\alpha<1$ blends the original and filtered data.
Applying the same right transformation to all constituent
blocks preserves their column alignment.
Substituting~\eqref{eq:blended_ivop} into the unified LMI gives
\[
F_{\mathrm{iv}}(P,\Sigma^d)
=
\Pi_{\alpha,\lambda}^{\top}
\widetilde F_a(P,\Sigma^d)
\Pi_{\alpha,\lambda},
\]
where $\widetilde F_a$ and $F_{\mathrm{iv}}$ are formed
from $\widetilde D_a$ and $D_{\mathrm{iv}}$, respectively.

For $0\leq\alpha<1$, $\Pi_{\alpha,\lambda}\succ0$,
so the congruence transformation is nonsingular and
preserves negative semidefiniteness in both directions.
Thus, preprocessing preserves the feasible set of the aligned
measured-data LMI in exact arithmetic and can improve
numerical conditioning, but does not establish a rigorous
guarantee for the underlying noise-free system.

For $\alpha=1$, the transformation may reduce the data rank.
Before applying RIW, we therefore check the numerical rank condition
$\operatorname{rank}(\widetilde Z_a\Pi_{\mathrm{iv},\lambda})=r$,
where $r$ is the required rank in
\eqref{eq:is-rank-condition}
or~\eqref{eq:data_driven_lag_selection},
depending on the formulation.
For noisy data, we retain the $r$ dominant singular directions
and apply RIW on the resulting subspace.

\section{CASE STUDY}
\label{sec:case-study}

This section evaluates the proposed framework on a modified
IEEE 39-bus system. The numerical study verifies behavioral
informativity under ambient conditions and examines how RIW
reduces the LMI dimension and improves computational efficiency,
and how IV preprocessing alleviates numerical ill-conditioning
and mitigates noise effects. The input-state and input-output
assessment results are presented for both noise-free and noisy
data, and the discretization error margins are analyzed.
A local controller update is used to demonstrate compositional
and decentralized stability reassessment.

\subsection{Test System and Preliminary Evaluation}

\begin{figure}[t]
    \centering
    \includegraphics[width=0.85\linewidth]{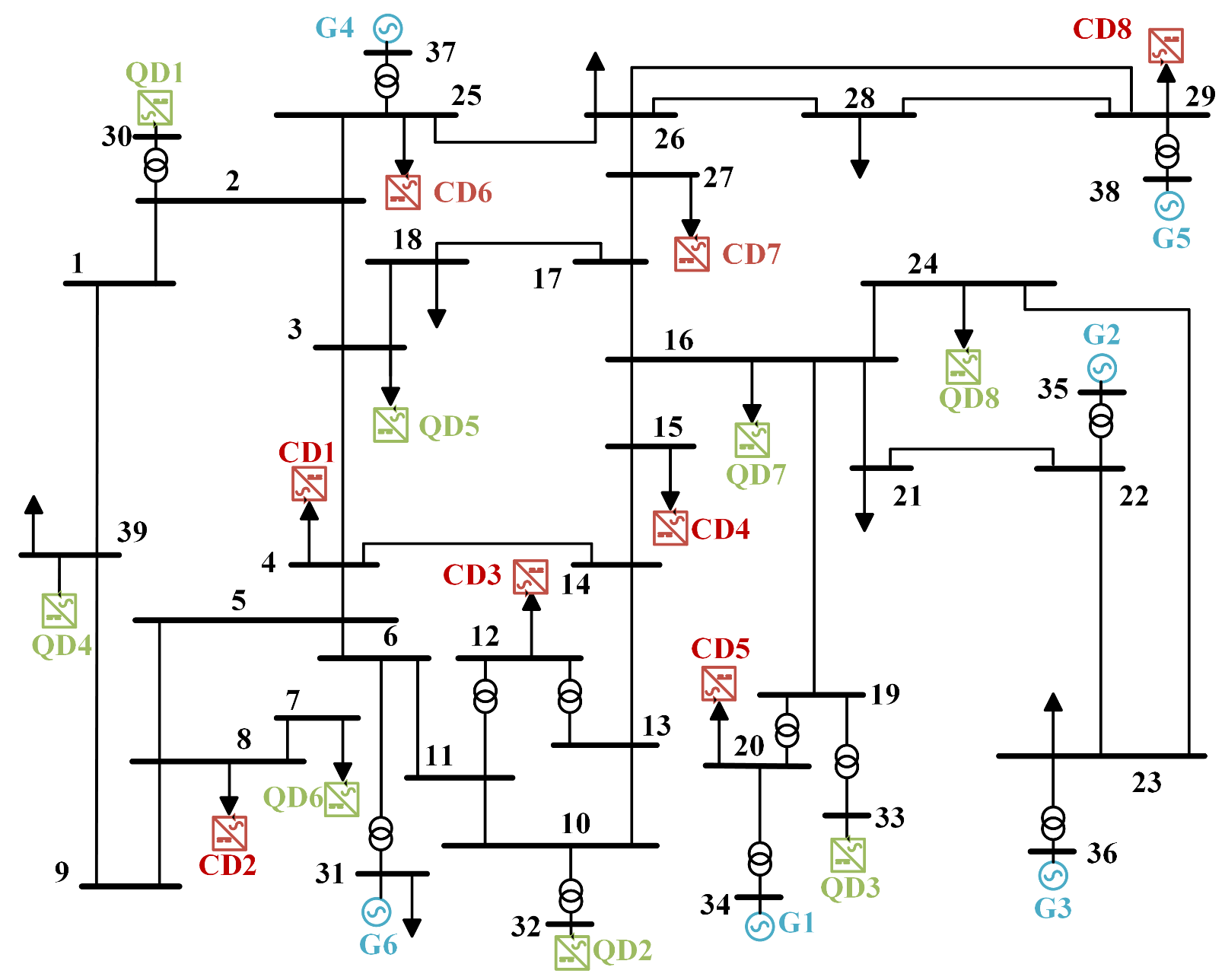}
    \caption{Schematic of the modified IEEE 39-bus power system.}
    \label{fig:39-bus}
\end{figure}

To evaluate the proposed method, we use a modified IEEE 39-bus system. Four synchronous generators (SGs)\cite{stegink2017unifying} are replaced by quadratic-droop-controlled inverters (QDs)\cite{simpson2016voltage}, while eight conventional-droop-controlled inverters (CDs)\cite{6987381} and four QDs are added at twelve load buses. The resulting system contains 6 SGs, 8 CDs, and 8 QDs, as shown in Fig.~\ref{fig:39-bus}. Buses without dynamic devices are eliminated by Kron reduction, with load fluctuations included as injection perturbations in the reduced network equations. In the simulations, we update the device states at each time step using a discretization of the continuous-time dynamic equations given in \cite{yang2019distributed} and the port inputs obtained by solving the network equations, with the local states and common network-port variables of the three device types summarized in Table~\ref{tab:device_variables} and the CD and QD droop time constants listed in Table~\ref{tab:inverter_time_constants}.

\begin{table}[htbp]
\centering
\footnotesize
\caption{LOCAL MEASUREMENT VARIABLES OF DEVICES}
\label{tab:device_variables}
\setlength{\tabcolsep}{3pt}
\renewcommand{\arraystretch}{1.1}
\begin{tabular*}{\columnwidth}{@{\extracolsep{\fill}} cccc @{}}
\hline
Device
& Local state $\boldsymbol{x}_i$
& Local input $\boldsymbol{u}_i$
& Local output $\boldsymbol{y}_i$ \\
\hline
SG &
$[\Delta\delta_i,\Delta\omega_i,\Delta E'_{qi}]^\top$ &
$[-\Delta P_i,-\Delta(Q_i/V_i)]^\top$ &
$[\Delta\theta_i,\Delta V_i]^\top$ \\

CD/QD &
$[\Delta\theta_i,\Delta V_i]^\top$ &
$[-\Delta P_i,-\Delta(Q_i/V_i)]^\top$ &
$[\Delta\theta_i,\Delta V_i]^\top$ \\
\hline
\end{tabular*}
\end{table}

\begin{table}[!t]
\centering
\footnotesize 
\caption{Time constants in the IEEE 39-bus system}
\label{tab:inverter_time_constants}
\resizebox{0.9\columnwidth}{!}{%
\begin{tabular}{c|cccccccc}
\hline
 & QD1 & QD2 & QD3 & QD4 & QD5 & QD6 & QD7 & QD8 \\
\hline
$\tau_{i1}$ (s) & 4.5 & 3.0 & 1.95 & 3.9 & 6.0 & 3.0 & 3.75 & 5.25 \\
$\tau_{i2}$ (s) & 7.0 & 7.5 & 8.0 & 8.2 & 8.5 & 6.5 & 9.2 & 9.6 \\
\hline
 & CD1 & CD2 & CD3 & CD4 & CD5 & CD6 & CD7 & CD8 \\
\hline
$\tau_{i1}$ (s) & 4.5 & 3.75 & 2.25 & 4.2 & 5.1 & 3.3 & 6 & 7.5 \\
$\tau_{i2}$ (s) & 8.1 & 9.0 & 9.5 & 9.3 & 10.0 & 9.5 & 7.0 & 6.5 \\
\hline
\end{tabular}
}
\end{table}

Under ambient conditions, stochastic load perturbations at six buses
are set according to Table~\ref{tab:load_perturbation}, and 1\% variations in renewable generation are applied at six CDs and six QDs. We check the resulting local trajectories for each device against the rank conditions in \eqref{eq:is-rank-condition} and
\eqref{eq:data_driven_lag_selection}. For SG, CD, and QD devices, respectively, the measured ranks are 5, 4, and 4 for the input-state formulation and 9, 6, and 6 for the input-output formulation. The lags calculated from~\eqref{eq:data_driven_lag_selection} are $l_d=2$, $1$, and $1$ for SG, CD, and QD devices, respectively. The sampling interval is $0.02$~s, and the simulation horizon is $60$~s.

\begin{table}[htbp]
\centering
\footnotesize
\caption{AMBIENT LOAD-FLUCTUATION LEVELS}
\label{tab:load_perturbation}
\begin{tabular*}{0.9\columnwidth}{@{\extracolsep{\fill}} c|cccccc @{}}
\hline
Load index & 18 & 21 & 23 & 26 & 28 & 31 \\
\hline
$\Delta P$ (p.u.) & 0.09 & 0.08 & 0.06 & 0.09 & 0.08 & 0.07 \\
$\Delta Q$ (p.u.) & 0.04 & 0.05 & 0.04 & 0.07 & 0.06 & 0.05 \\
\hline
\end{tabular*}
\end{table}

\begin{figure}[htbp]
    \centering

    \begin{minipage}{0.49\linewidth}
        \centering
        \includegraphics[width=\linewidth]{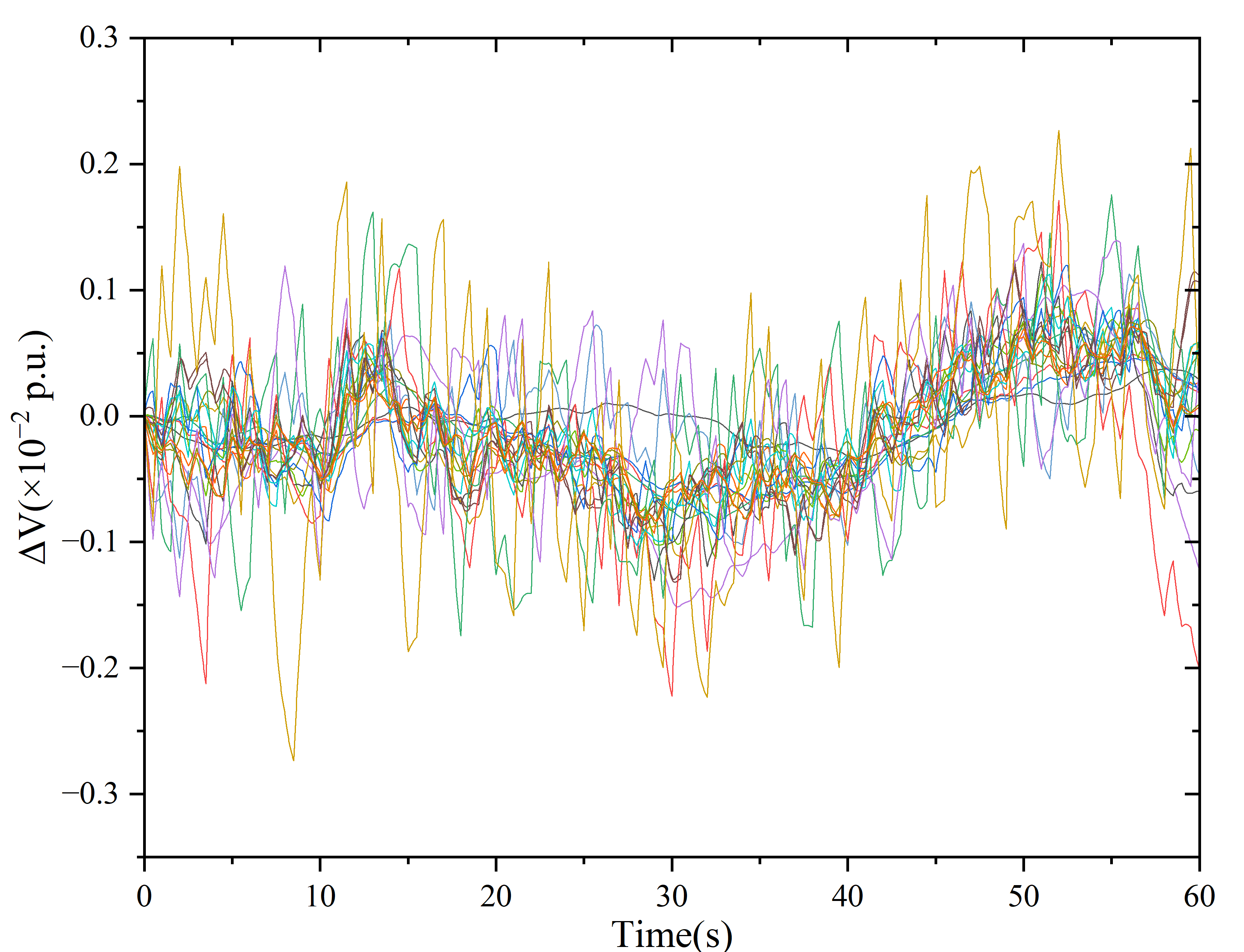}

    \end{minipage}
    \hfill 
    \begin{minipage}{0.49\linewidth}
        \centering
        \includegraphics[width=\linewidth]{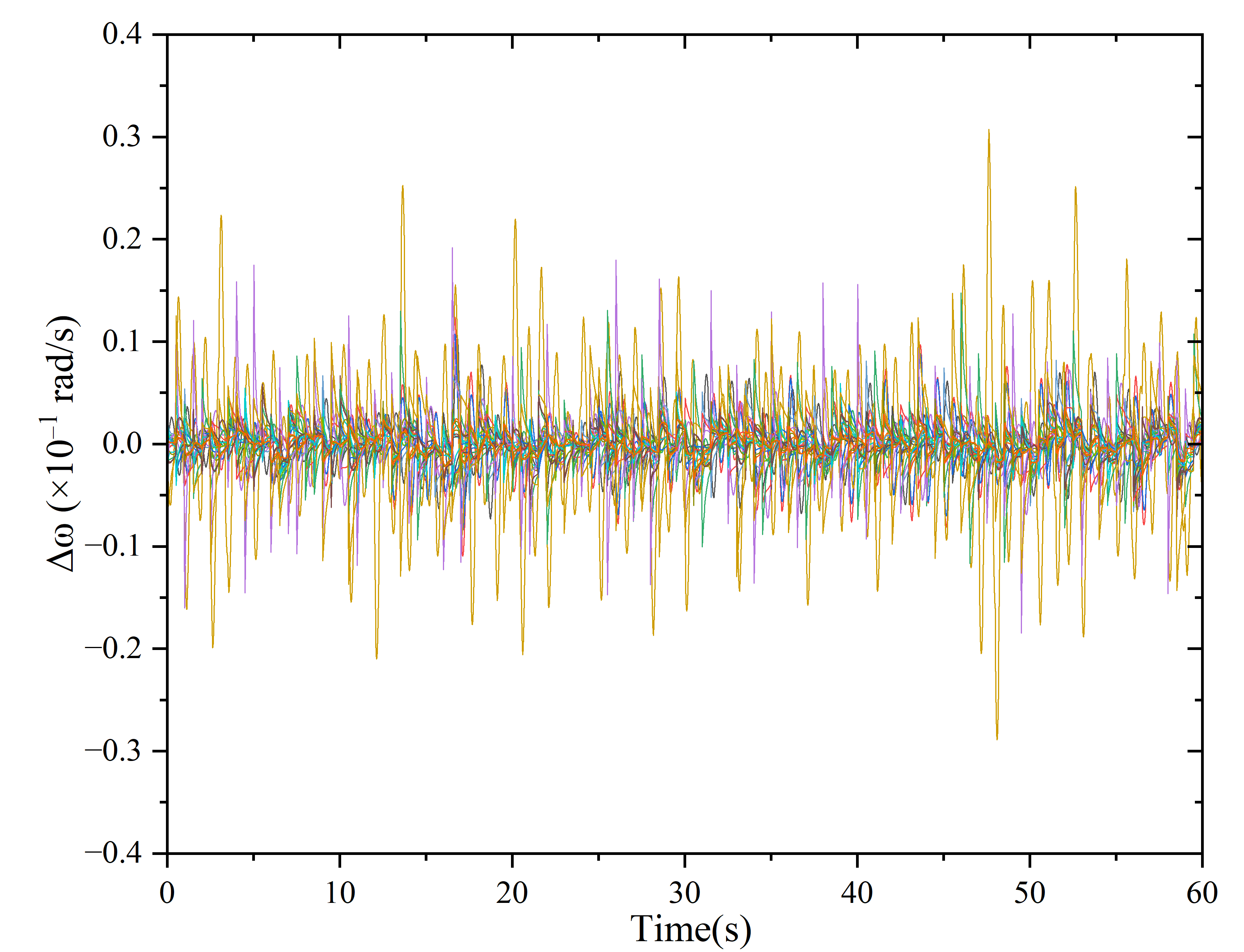}

    \end{minipage}

    \caption{Time-domain responses under ambient conditions. (A) Voltage deviation $\Delta V$. (B) Angular-frequency deviation $\Delta\omega$.}
    \label{fig:time_domain_response}
    
\end{figure}

Under the ambient conditions, the resulting voltage and angular-frequency responses are shown in Fig.~\ref{fig:time_domain_response}. The voltage deviations remain within approximately $0.25\times10^{-2}$~p.u., while the angular-frequency deviations remain within approximately $0.3\times10^{-1}$~rad/s. The rank results confirm behavioral informativity, while the response magnitudes show that the system remains in a small neighborhood of the equilibrium. Together, these results support the small-signal LTI approximation adopted in Section~\ref{sec:divice-network} and the subsequent input-state and input-output ODP studies.

RIW compresses the data-driven LMI from the original data-column space onto the numerical behavioral subspace. As shown by the representative single-device tests in Table~\ref{tab:rinv_whitening_cost}, the LMI dimension decreases from more than $10^3$ to $4$--$9$. The SG and QD input-output LMIs run out of memory, whereas the whitened SDPs remain numerically tractable.

\begin{table}[t]
\centering
\caption{Computational Efficiency of RIW}
\label{tab:rinv_whitening_cost}
\resizebox{0.9\columnwidth}{!}{%
\begin{tabular}{@{} l c c c c @{}} 
\hline
\multirow{2}{*}{Case} & Original & Reduced & \multicolumn{2}{c}{Time (s)} \\
\cline{4-5} 
 & LMI Dim. & LMI Dim. & w/o RIW & w/ RIW \\
\hline
Input--state SG  & 1448 & 5 & 295.046 & 2.78 \\
Input--state QD  & 1448 & 4 & 251.235 & 1.146 \\
Input--state CD  & 1448 & 4 & 227.002 & 1.436 \\
Input--output SG & 1433 & 9 & Out of memory & 1.391 \\
Input--output QD & 1434 & 6 & Out of memory & 2.379 \\
Input--output CD & 1434 & 6 & 400.273 & 1.354 \\
\hline
\end{tabular}%
}
\end{table}

\begin{figure}[t]
    \centering
    \includegraphics[width=0.65\linewidth]{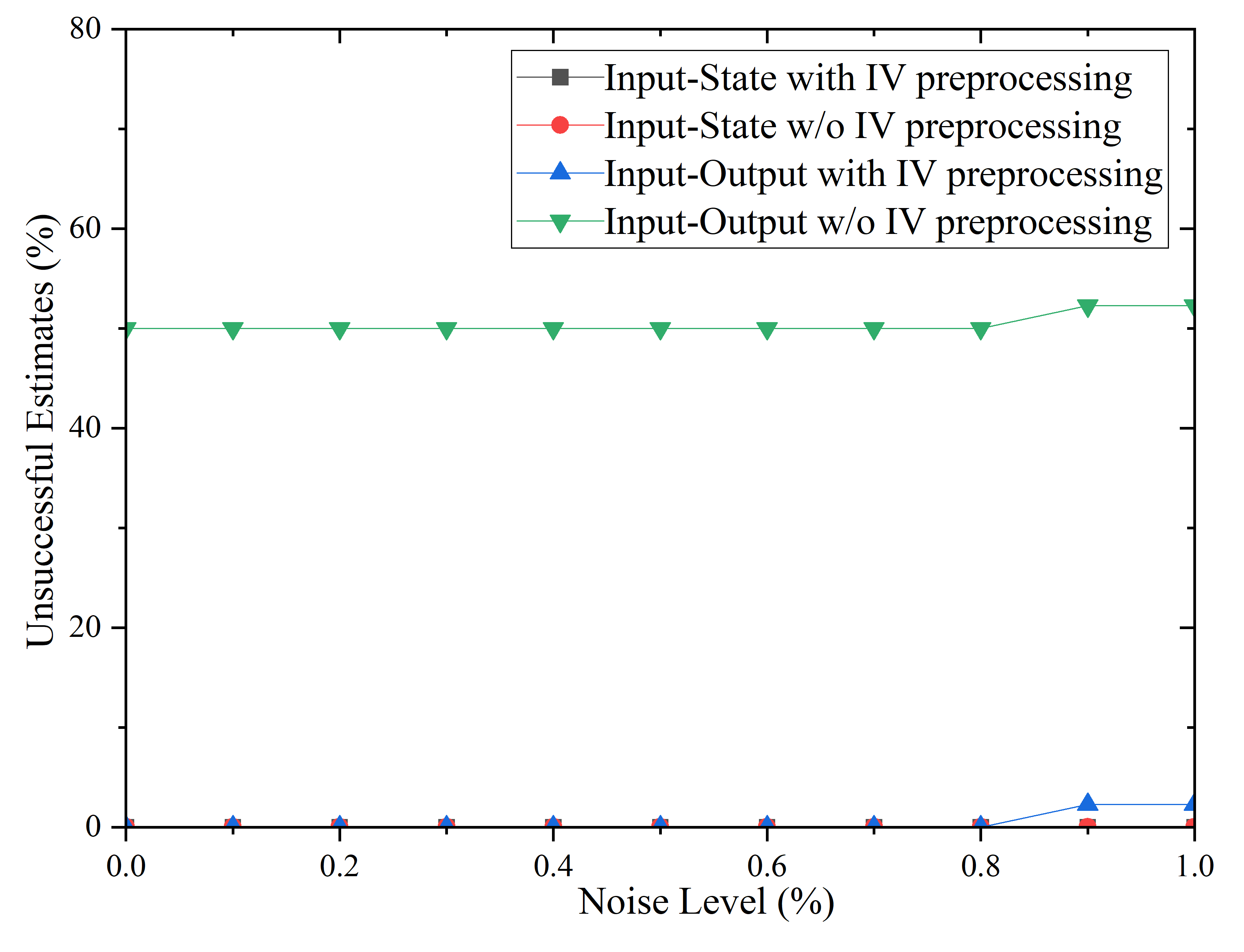}
    \caption{Fraction of unsuccessful estimates with and without IV preprocessing.}
    \label{fig:abnormal_rate}
\end{figure}

At each measurement-noise level, channel-wise relative noise is applied to the
$X$, $U$ and $Y$ data entering the LMI; the 1\% case is used as a
PMU-scale benchmark motivated by \cite{7052413}. Fig.~\ref{fig:abnormal_rate}
compares the fraction of unsuccessful estimates for the input-state and
input-output formulations. For each noise level, the fraction is computed
over device-window pairs: an estimate is classified as unsuccessful if
the SDP fails or if the lower bound $s_i$ falls outside the
$s_i^{ref}\pm0.2$ band, where $s_i^{ref}$ denotes the model-based reference.
Without IV preprocessing, this fraction is approximately 50\% for the input-output formulation even without added measurement noise and increases slightly with noise. In contrast, IV preprocessing reduces this rate to about 2\% at 1\% noise, effectively alleviating numerical ill-conditioning.

\subsection{Estimation Settings and Discretization Error Margins}
\label{sec:ivop_comparison}

To evaluate the proposed method under different controller settings,
we configure all device controllers according to a theoretical ODP
reference $\sigma$ and consider seven settings with $\sigma$
decreasing from $2.8146$ to $2.0146$.
At each point, a full symmetric matrix $\Sigma_i^d$ and its lower
bound $s_i$ are estimated for every device. The decentralized and
compositional margins are defined as
$\mu_{dec}^d:=\min_i(s_i-\sigma_{net})$ and
$\mu_{com}^d:=\lambda_{\min}(H_{net}+D_{\Sigma^d})$,
respectively.

For each device, the $60\,\mathrm{s}$ trajectory is first processed
by a first-order low-pass filter with a $0.02\,\mathrm{s}$ time
constant and then divided into two $29\,\mathrm{s}$ windows,
whose matrix estimates are averaged. RIW retains the leading $r$ singular directions, where $r$
is the required behavioral rank in
\eqref{eq:is-rank-condition} or
\eqref{eq:data_driven_lag_selection},
depending on the formulation.
IV preprocessing precedes RIW and the SDP.
For the input-state case, $\tau=2$, $\lambda=10^{-5}$, and
$\alpha=0.6$ for SGs and $1$ for CDs and QDs.
SG estimation is numerically successful for $\alpha\in[0.5,0.8]$,
whereas $\alpha=1$ attenuates a weak behavioral direction;
for CDs and QDs, $\alpha=1$ satisfies the rank check.

For the input-output formulation, the extended state is
constructed from local input-output trajectories using the
data-selected $l_d$. The same filtering, windowing,
IV--RIW sequence, and SDP are used, with alignment accounting
for both $l_d$ and $\tau$. The parameters $\lambda$ and
$\alpha$ remain unchanged, whereas $\tau=15$
($0.3\,\mathrm{s}$) is used to reduce the structured noise
correlation introduced by overlapping historical samples.

To assess the discretization error margins, an offline
resolution-refinement study is conducted at
$h_0=0.02\,\mathrm{s}$ and two calibration points,
$\sigma_{\mathit{cal}}\in\{2.8146,2.5479\}$,
to define an empirical decision buffer $\varepsilon_{\mathit{fs}}$.
The high-resolution trajectory is sampled at
$h_0$, $h_0/2$, and $h_0/4$, with the IV lag adjusted
to preserve the physical delay.
For the input-output formulation, $l_d(h)$ is reselected
at each resolution.

At each calibration point, we compute the difference between
the largest and smallest values of each stability margin
obtained at the three sampling resolutions.
Taking the largest difference over the two calibration points
and both the compositional and decentralized margins gives
$0.02$ for IS and $0.04$ for IO. We therefore adopt $\varepsilon_{\mathit{fs}}=0.04$ as a common empirical decision buffer for the following stability assessments. Each stability test is considered satisfied only when its
estimated margin exceeds $\varepsilon_{\mathit{fs}}$.

\subsection{Data-Driven ODP Estimation and Stability Assessment}

\begin{figure}[t]
    \centering
    \includegraphics[width=0.9\linewidth]{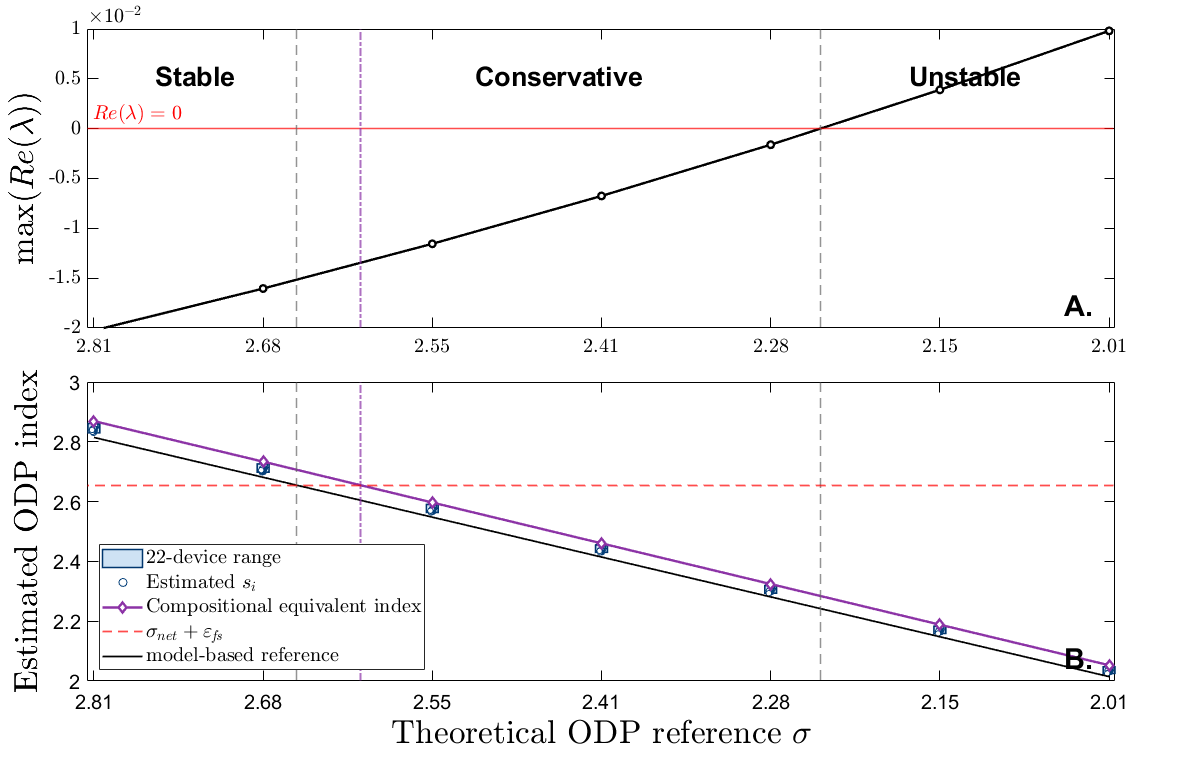}
    \caption{Input-state stability assessment without added
    measurement noise.
    (A) Maximum real part of the analytical eigenvalues.
    (B) Estimated lower bounds $s_i$, their 22-device range,
    the model-based reference, and the decision threshold
$\sigma_{net}+\varepsilon_{\mathit{fs}}$.
    The purple curve denotes the compositional equivalent index
$s_{com}^{eq}
:=\sigma_{net}+\mu_{com}^{d}$;
the buffer-adjusted compositional test requires
$s_{com}^{eq}>
\sigma_{net}+\varepsilon_{\mathit{fs}}$.}
    \label{fig:input_state_noisefree}
\end{figure}

\begin{figure}[t]
    \centering
    \includegraphics[width=0.9\linewidth]{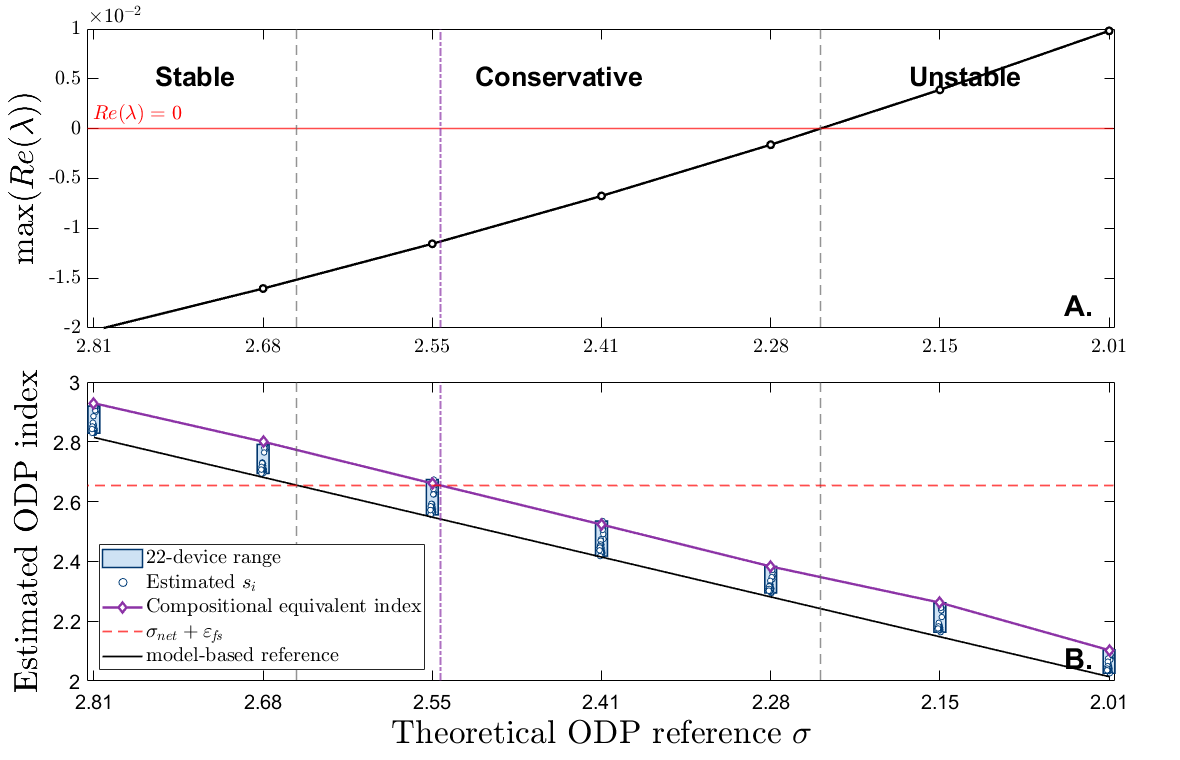}
    \caption{Input-output stability assessment without added
    measurement noise.}
    \label{fig:input_output_noisefree}
\end{figure}

Without added measurement noise, the estimated lower bounds
for both the input-state and input-output formulations remain
close to the model-based reference, as shown in
Figs.~\ref{fig:input_state_noisefree}
and~\ref{fig:input_output_noisefree}.
The upper panels locate the analytical stability boundary of system~\eqref{eq:interconnected_system} at $\sigma=2.24$.
In the lower panels, the black model-based reference
crosses $\sigma_{net}+\varepsilon_{\mathit{fs}}\approx2.6546$,
as marked by the left gray vertical line.
Since the estimated $s_i$ closely follow this reference,
no separate decentralized boundary is shown.

\begin{figure}[t]
    \centering
    \includegraphics[width=0.9\linewidth]{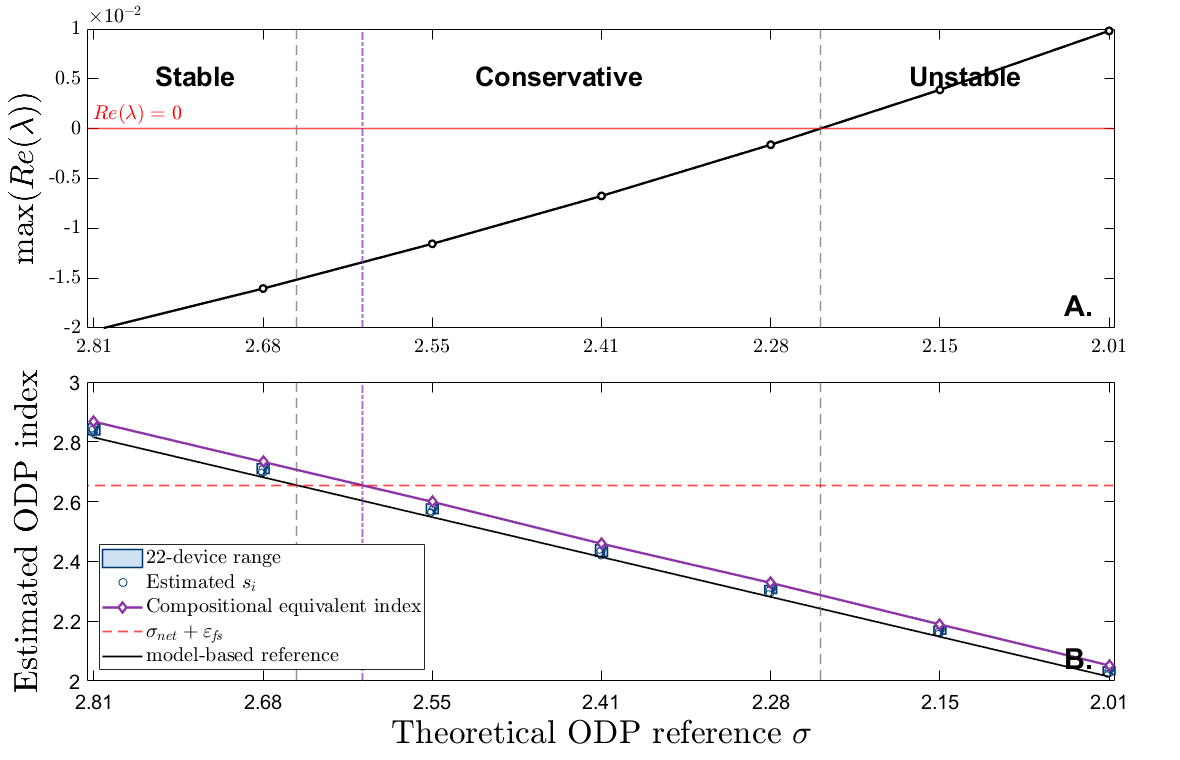}
    \caption{Input-state stability assessment under $1\%$
    measurement noise.}
    \label{fig:input_state_noisy}
\end{figure}

\begin{figure}[t]
    \centering
    \includegraphics[width=0.9\linewidth]{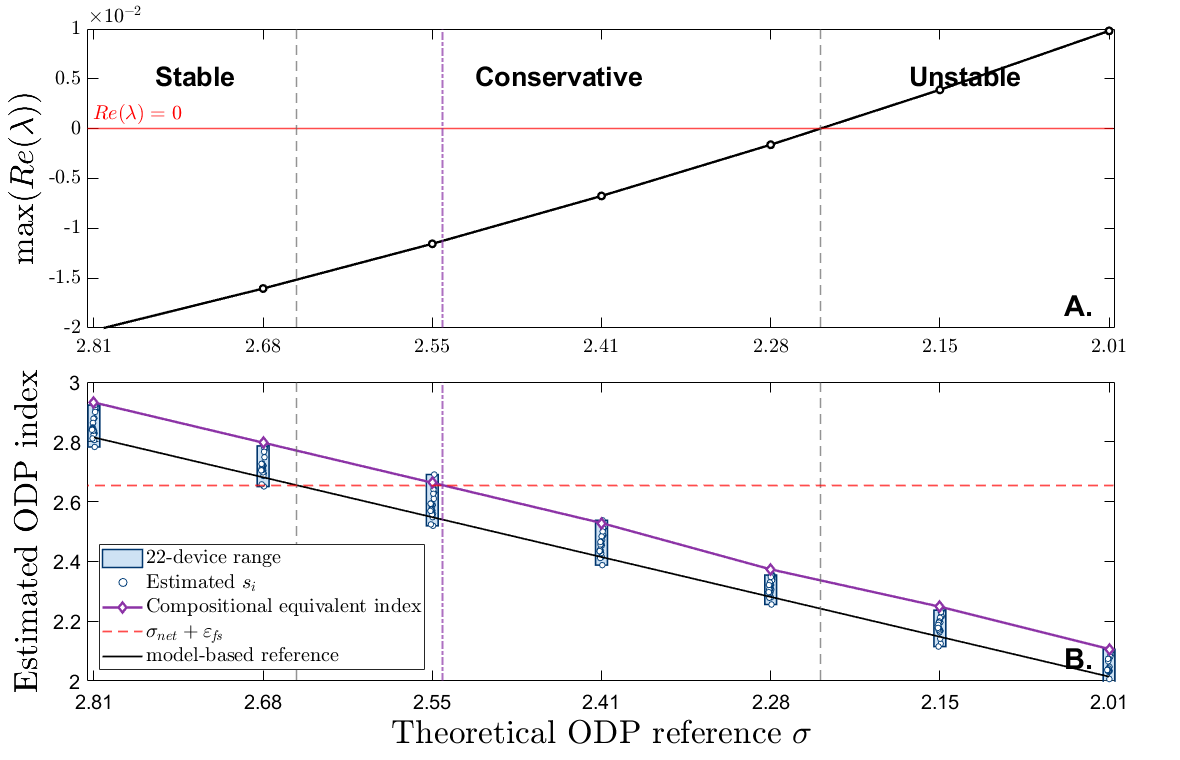}
    \caption{Input-output stability assessment under $1\%$
    measurement noise.}
    \label{fig:input_output_noisy}
\end{figure}

Under $1\%$ measurement noise, the estimated lower bounds
for both formulations remain close to the model-based
reference, as shown in Figs.~\ref{fig:input_state_noisy}
and~\ref{fig:input_output_noisy}.

For both noise-free data and data with $1\%$ measurement
noise, the same scan points satisfy the stability tests,
and all lie in the model-based asymptotically stable region.
For IS, both tests are satisfied at the first two scan points.
For IO, the compositional margin at $\sigma=2.5479$
remains above $\varepsilon_{\mathit{fs}}$, whereas the
decentralized margin does not, as summarized in
Table~\ref{tab:finite_sampling_assessment}.
These results, together with the purple boundary extending
farther into the conservative region, demonstrate the reduced
conservatism obtained by retaining the complete matrix structure.

\begin{table}[t]
    \centering
    \footnotesize
    \caption{Sampling-resolution differences and scan points
    satisfying the stability tests with
    $\varepsilon_{\mathit{fs}}=0.04$.}
    \label{tab:finite_sampling_assessment}
    \setlength{\tabcolsep}{3pt}
    \begin{tabular*}{0.9\columnwidth}
    {@{\extracolsep{\fill}}lccc@{}}
        \hline
        Setting & Max. difference & COM & DEC \\
        \hline
        IS & $0.02$
        & $2.8146,\,2.6813$
        & $2.8146,\,2.6813$ \\
        IO & $0.04$
        & $2.8146,\,2.6813,\,2.5479$
        & $2.8146,\,2.6813$ \\
        \hline
    \end{tabular*}
\end{table}

\subsection{Modular Stability Reassessment After Local Controller Updates}

The preceding studies evaluate the discretization effect and show close agreement with model-based references under the tested noise levels. We next examine how the matrix estimates and lower bounds support compositional and decentralized reassessment under a local controller update. The same empirical decision buffer $\varepsilon_{fs}=0.04$ is used throughout the update scan.

SG5 is selected as the updated device, while the theoretical ODP references of the remaining 21 devices are fixed at $2.7$. Its control parameters $K_I(5)$ and $K_E(5)$ are jointly varied to adjust the theoretical reference $\sigma_{5,ref}$. In the adopted SG model \cite{yang2019distributed}, these gains multiply feedback deviations that vanish at the prescribed equilibrium; thus, their variation with fixed setpoints preserves the equilibrium power flow. Because the topology, network parameters, and equilibrium
power flow remain unchanged, $H_{net}$ and $\sigma_{net}$ remain fixed. At each parameter value, only $\Sigma_5^d$ and $s_5$ are updated from the SG5 trajectory, while the other 21 local estimates are retained.

\begin{figure}[t]
\centering
\includegraphics[width=\linewidth]{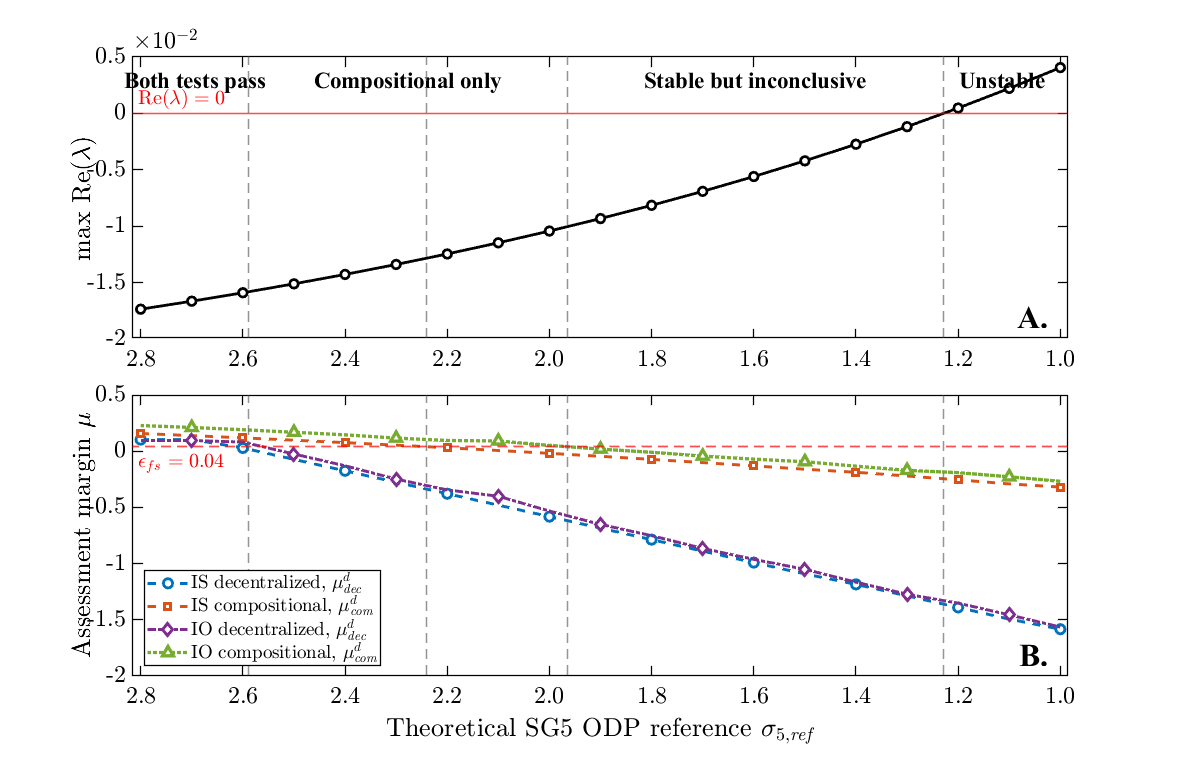}
\caption{Data-driven compositional and decentralized stability assessment under an equilibrium-preserving local controller update of SG5. Region labels summarize the overall progression, and setting-specific boundaries and conservatism reductions are listed in Table~\ref{tab:local_update_boundaries}.}
\label{fig:local_update_assessment}
\end{figure}

As $\sigma_{5,ref}$ decreases in Fig.~\ref{fig:local_update_assessment}, the decentralized test fails first, while the compositional condition remains satisfied over a wider range. Further reduction leaves the analytical system stable but the data-driven assessment inconclusive before the analytical system becomes unstable at $\sigma_{5,ref}=1.2287$. The corresponding assessment boundaries are listed in Table~\ref{tab:local_update_boundaries}.

\begin{table}[t]
\centering
\footnotesize
\caption{ASSESSMENT BOUNDARIES FOR THE SG5 UPDATE}
\label{tab:local_update_boundaries}
\begin{tabular*}{\columnwidth}
{@{\extracolsep{\fill}}lcccc@{}}
\hline
Data & Decentralized & Compositional & Analytical & Reduction \\
\hline
Input--state  & 2.615 & 2.241 & 1.2287 & 26.98\% \\
Input--output & 2.561 & 1.965 & 1.2287 & 44.74\% \\
\hline
\end{tabular*}
\end{table}

As summarized in Table~\ref{tab:local_update_boundaries}, the
compositional test reduces the decentralized-to-analytical conservative
interval by $26.98\%$ and $44.74\%$ for the input-state and
input-output formulations, respectively. This difference arises from
their distinct matrix estimates when combined with $H_{net}$.

\begin{figure}[t]
\centering
\includegraphics[width=\linewidth]{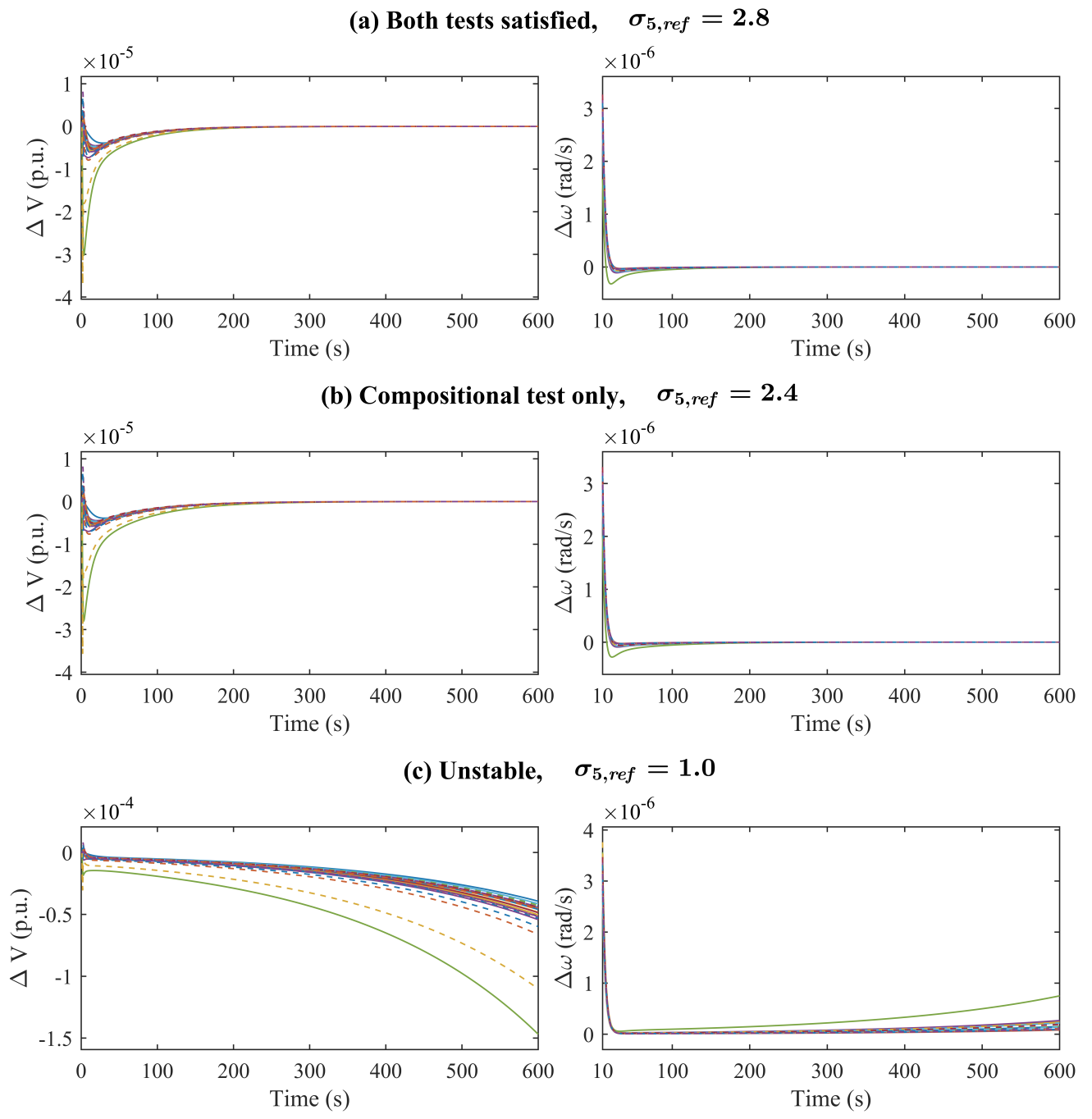}
\caption{Time-domain responses under three SG5 controller
    settings. Left: voltage deviations. Right: angular-frequency
    deviations shown over $t\in[10,600]\,\mathrm{s}$.
    Each panel shows the responses of all 22 devices.}
\label{fig:sg5_time_domain}
\end{figure}

To further evaluate stability reassessment after local controller
updates, three SG5 settings,
$\sigma_{5,\mathit{ref}}\in\{2.8,2.4,1.0\}$, are selected to
represent the cases where both stability tests are satisfied,
only the compositional test is satisfied, and the analytical
system is unstable, respectively.
The control parameters of the remaining 21 devices are unchanged.
The same small active-power load pulse is applied at bus 28
over $t\in[1,2]\,\mathrm{s}$, with an amplitude equal to 1\%
of its equilibrium active load, to compare the responses around
the same equilibrium after the pulse is removed.
Fig.~\ref{fig:sg5_time_domain} shows the voltage and
angular-frequency deviations of all 22 devices.
At $\sigma_{5,\mathit{ref}}=2.8$ and $2.4$, the responses decay
toward the equilibrium.
In particular, the convergent responses at
$\sigma_{5,\mathit{ref}}=2.4$, where only the compositional test
is satisfied, further demonstrate the reduced conservatism
of the compositional assessment.
At $\sigma_{5,\mathit{ref}}=1.0$, the responses grow after
the initial transient.

Together with the preceding assessment results, these
time-domain responses support modular and scalable stability
reassessment after local controller updates.

\section{CONCLUSION}
\label{sec:conclusion}

In this paper, a compositional data-driven small-signal stability assessment
framework based on ODP is proposed for power systems under
ambient conditions. The proposed method estimates local ODP
matrices and scalar indices of heterogeneous devices from
input-state or input-output trajectories and uses them in both compositional and decentralized
Lyapunov-stability tests.
To address the computational and numerical challenges of
trajectory-based LMIs, we introduce RIW to reduce the LMI
dimension and improve computational efficiency, and use IV
preprocessing to alleviate numerical ill-conditioning and
mitigate noise effects.

Future work will investigate rigorous stability guarantees under measurement noise, real-world validation, and data-driven controller synthesis.

\appendices
\section{}
\label{app:proofs}
\newcommand{\proofheading}[2]{%
  \refstepcounter{subsection}%
  \subsection*{\thesubsection.\ #1}%
  \label{#2}}
\proofheading{Proof of Lemma~\ref{lem:compositional-stability}}{app:proof-compositional}

\begin{proof}
Consider the Lyapunov function candidate
\[
\begin{aligned}
W(x)
&=
\sum_{i\in\mathcal V}\bigl(S_i(x_i)-S_i(0)\bigr)
+W_N(y)+\frac12 y^\top D_\Sigma y\\
&=
\sum_{i\in\mathcal V}\bigl(S_i(x_i)-S_i(0)\bigr)
+\frac12 y^\top(H_{\mathrm{net}}+D_\Sigma)y.
\end{aligned}
\]
Each local storage function $S_i$ has a strict local minimum
at $x_i=0$, and $H_{\mathrm{net}}+D_\Sigma\succ0$.
Thus, $W(0)=0$, and $W$ is positive definite in a neighborhood
of the origin.

The network potential and the conductance-related mapping are
expressed in the incremental coordinates $y$ defined in the
main text, with the equilibrium corresponding to $y=0$.
We follow \cite{yang2019distributed} for the power-flow
decomposition and the numerical energy function construction.
For each bus $i$, the power-flow decomposition gives
\begin{equation}
\label{eq:app-a-power-flow-decomposition}
\begin{aligned}
P_i=\frac{\partial\widetilde W_B(y)}{\partial\theta_i}+\phi_{pi}(y),\quad \frac{Q_i}{V_i}=\frac{\partial\widetilde W_B(y)}{\partial V_i}+\phi_{qi}(y).
\end{aligned}
\end{equation}
Subtracting the corresponding equilibrium values in
\eqref{eq:app-a-power-flow-decomposition} and using the
definition of the incremental port input yields
\begin{equation}
\label{eq:app-a-incremental-port-mapping}
-u=
\nabla\widetilde W_B(y)-\nabla\widetilde W_B(0)
+\Phi(y)-\Phi(0),
\end{equation}
where $\widetilde W_B$ is the susceptance-related network
potential and
$\Phi(y):=\operatorname{col}_{i\in\mathcal V}
\bigl(\phi_{pi}(y),\phi_{qi}(y)\bigr)$
collects the conductance-related terms given by
\[
\begin{aligned}
\phi_{pi}(y)
&=
G_{ii}V_i^2+
\sum_{j\in\mathcal N_i}
G_{ij}V_iV_j\cos\theta_{ij},\\
\phi_{qi}(y)
&=
\sum_{j\in\mathcal N_i}
G_{ij}V_j\sin\theta_{ij}.
\end{aligned}
\]
Here, $\theta_i$ and $V_i$ are the actual voltage angle and
magnitude corresponding to the incremental coordinates $y$.

Let $H_B:=\nabla^2\widetilde W_B(0)$ and
$A_G:=\left.\partial\Phi/\partial y\right|_{y=0}$.
Differentiating \eqref{eq:app-a-incremental-port-mapping}
at the equilibrium gives $J_{\mathrm{net}}=-H_B-A_G$.
In the numerical energy function construction, we define
\begin{equation}
\label{eq:app-a-conductance-energy}
W_G(y):=\frac12 y^\top A_Gy
\end{equation}
and adopt the approximation
\begin{equation}
\label{eq:app-a-energy-approximation}
\nabla W_G(y)\approx\Phi(y)-\Phi(0).
\end{equation}
For the linearized network, retaining the quadratic term
of the susceptance-related incremental potential and adding
$W_G(y)$ from \eqref{eq:app-a-conductance-energy} yields
\[
\begin{aligned}
W_N(y)
&=\frac12 y^\top H_By+W_G(y)\\
&=\frac12 y^\top
\left[H_B+\frac12(A_G+A_G^\top)\right]y=\frac12 y^\top H_{net}y,\\
\end{aligned}
\]
where
$H_{net}
=
H_B+\frac12(A_G+A_G^\top)
=
-\frac12(J_{net}+J_{net}^\top),
$
consistent with its definition in the main text.

The first-order conductance-related term in
\eqref{eq:app-a-incremental-port-mapping} is $A_Gy$.
Under \eqref{eq:app-a-energy-approximation}, we replace
this term with $\nabla W_G(y)$. In the following calculations,
$u$ denotes the device port input under this approximation
and satisfies
\[
-u=H_By+\nabla W_G(y)=H_{net}y.
\]
Thus, by the chain rule,
\[
\begin{aligned}
\dot W_N(y)
&=\bigl(H_By+\nabla W_G(y)\bigr)^\top\dot y=-u^\top\dot y.
\end{aligned}
\]

The local ODP inequality for each device is
$\dot S_i(x_i)\le
u_i^\top\dot y_i-y_i^\top\Sigma_i\dot y_i$.
Summing over all devices yields
\[
\sum_{i\in\mathcal V}\dot S_i(x_i)
\le u^\top\dot y-y^\top D_\Sigma\dot y.
\]
Since $D_\Sigma$ is symmetric and fixed at the given
operating point, the derivative of
$\frac12 y^\top D_\Sigma y$ is $y^\top D_\Sigma\dot y$.
Consequently,
\[
\begin{aligned}
\dot W(x)
&=
\sum_{i\in\mathcal V}\dot S_i(x_i)
+\dot W_N(y)+y^\top D_\Sigma\dot y\\
&\le
u^\top\dot y-y^\top D_\Sigma\dot y
-u^\top\dot y+y^\top D_\Sigma\dot y=0.
\end{aligned}
\]
Thus, $W$ is locally positive definite and nonincreasing
along the trajectories of the interconnected power system.
The equilibrium is therefore Lyapunov stable under the
adopted numerical energy function approximation.
\end{proof}
\begin{remark}
The above proof addresses lossy networks and follows the numerical
energy function approximation in~\cite{yang2019distributed}
to treat the conductance-related nonconservative terms.
Numerical energy functions are a common approach to handling
network losses in power-system direct methods, with a long
history of research and
application~\cite{chiang1995direct,guo2025efficient}.
Moreover, high-voltage transmission lines are typically dominated
by reactance, and neglecting their relatively small resistive
losses to adopt a lossless network model is also common in grid
dynamic analysis; the corresponding modeling rationale can be
found in~\cite{schiffer2024control}.
When the network is lossless ($G=0$), the conductance-related
terms $\Phi$ and $W_G$ vanish, and the above network energy
relation holds exactly for the corresponding linearized model,
without requiring the numerical energy function approximation
for the conductance-related terms.
\end{remark} 

\proofheading{Proof of Corollary~\ref{cor:decentralized-stability}}{app:proof-decentralized}

\begin{proof}
Let $D_s:=\operatorname{blkdiag}(s_i I_{m_i})$.
By definition, $H_{net}\succeq-\sigma_{net}I$.
The inequalities $s_i>\sigma_{net}$ for all $i\in\mathcal V$
imply $D_s\succ\sigma_{net}I$.
Since $D_\Sigma\succeq D_s$, it follows that
\[
H_{net}+D_\Sigma\succeq H_{net}+D_s\succ0.
\]
The result follows from Lemma~\ref{lem:compositional-stability}.
\end{proof}

\proofheading{Proof of Lemma~\ref{lem:ct-lti}}{app:proof-ct-lmi}
\begin{proof}
Let $z=[x^\top\;u^\top]^\top$ and consider
$S(x)=x^\top Px$ with $P=P^\top\succ0$.
Along the trajectories of the continuous-time LTI system,
\begin{align*}
\dot{S}(x)
&=(Ax+Bu)^\top Px
  +x^\top P(Ax+Bu)\\
&=x^\top(A^\top P+PA)x
  +2x^\top PBu\\
&=z^\top M_1z.
\end{align*}
Since $y=Cx$ and
$\dot{y}=CAx+CBu$, the ODP supply rate satisfies
\begin{align*}
u^\top\dot{y}-y^\top\Sigma\dot{y}
&=(u-\Sigma Cx)^\top
  (CAx+CBu)\\
&=z^\top M_2z=\frac{1}{2}z^\top
  (M_2+M_2^\top)z,
\end{align*}
where the last equality follows because a scalar quadratic
form is unchanged by symmetrizing its matrix.
Therefore, the ODP inequality~\eqref{eq:matrix_odp} holds if and only if
the following inequality holds for all $z$:
\[
z^\top\left[
M_1-\frac{1}{2}
(M_2+M_2^\top)
\right]z\leq0.
\]
This is equivalent to the stated LMI~\eqref{eq:ct-lti}.
Moreover, $P\succ0$ ensures that $S(x)$ has a strict minimum
at the origin. Hence, the LMI is necessary and sufficient
for the existence of the required quadratic storage function.
\end{proof}

\proofheading{Proof of Theorem~\ref{thm:dt-odp-lmi}}{app:proof-dt-lmi}
\begin{proof}
Let $z_k=[x_k^\top\;u_k^\top]^\top$ and consider
$S(x_k)=x_k^\top Px_k$ with $P=P^\top\succ0$.
Using $x_{k+1}=\bar{A}x_k+\bar{B}u_k$, the storage increment satisfies
\begin{align*}
S(x_{k+1})-S(x_k)
&=(\bar{A}x_k+\bar{B}u_k)^\top P(\bar{A}x_k+\bar{B}u_k)
  -x_k^\top Px_k\\
&=z_k^\top\bar{M}_1z_k.
\end{align*}
Moreover, $y_k=Cx_k$ gives
$\Delta y_k=C(\bar{A}-I)x_k+C\bar{B}u_k$. Hence,
\begin{align*}
(u_k-\Sigma^d y_k)^\top\Delta y_k
&=(u_k-\Sigma^d Cx_k)^\top
  \bigl(C(\bar{A}-I)x_k+C\bar{B}u_k\bigr)\\
&=z_k^\top\bar{M}_2z_k\\
&=\frac{1}{2}z_k^\top(\bar{M}_2+\bar{M}_2^\top)z_k.
\end{align*}
For this quadratic storage function, the dissipation inequality
\eqref{eq:dt-odp-dissipation} is therefore equivalent to
\[
z_k^\top\left[
\bar{M}_1-\frac{1}{2}(\bar{M}_2+\bar{M}_2^\top)
\right]z_k\leq0.
\]

If the system satisfies ODP($\Sigma^d$) with a quadratic
storage function, then its matrix $P=P^\top\succ0$ satisfies
the above inequality.
Since $x_0$ and $u_0$ are arbitrary, the quadratic form is
nonpositive for all $z_0\in\mathbb{R}^{n+m}$, yielding
the LMI~\eqref{eq:dt-odp-lmi}.

Conversely, if \eqref{eq:dt-odp-lmi} holds for some
$P=P^\top\succ0$, the above identities show that
$S(x)=x^\top Px$ satisfies \eqref{eq:dt-odp-dissipation}
along every system trajectory. Thus, the system satisfies
ODP($\Sigma^d$).
\end{proof}

\proofheading{Proof of the Discretization Error-Margin Relation}{app:proof-discretization}

\begin{proof}
Fix $P\succ0$ and a symmetric matrix $\Sigma^d$.
By \eqref{eq:exact-discretization}, as $h\to0$,
$\bar{A}=I+hA+O(h^2)$ and $\bar{B}=hB+O(h^2)$.

Substituting these expansions into the matrices $\bar{M}_1$ and
$\bar{M}_2$ of the discrete-time ODP condition~\eqref{eq:dt-odp-lmi}
yields
\[
\bar{M}_1=hM_1+O(h^2),
\qquad
\bar{M}_2=hM_2+O(h^2),
\]
where $M_1$ and $M_2$ are the corresponding matrices in the
continuous-time ODP condition~\eqref{eq:ct-lti}, evaluated at
$\Sigma=\Sigma^d$. Therefore,
\[
\begin{aligned}
R_d(P,\Sigma^d;h)
&=\bar{M}_1-\tfrac12(\bar{M}_2+\bar{M}_2^\top)\\
&=h\!\left[M_1-\tfrac12(M_2+M_2^\top)\right]+O(h^2)\\
&=hR_c(P,\Sigma^d)+O(h^2).
\end{aligned}
\]
Dividing both sides by $h$ gives
$h^{-1}R_d(P,\Sigma^d;h)=R_c(P,\Sigma^d)+E_h(P,\Sigma^d)$,
where $\|E_h(P,\Sigma^d)\|_2=O(h)$.
\end{proof}

\proofheading{Proof of Theorem~\ref{thm:main}}{app:proof-main}
\begin{proof}
For 1), the assumed discretization error margins imply
\[
D_{\Sigma}
=
\operatorname{blkdiag}(\Sigma_i)
\succeq
\operatorname{blkdiag}
\bigl(\Sigma_i^d-\varepsilon_i I_{m_i}\bigr)
=
D_{\Sigma^d}-D_{\varepsilon}.
\]
Therefore, under the condition
$H_{net}+D_{\Sigma^d}\succ D_{\varepsilon}$,
we obtain
\[
H_{net}+D_{\Sigma}
\succeq
H_{net}+D_{\Sigma^d}-D_{\varepsilon}
\succ0.
\]
Thus, the compositional stability condition is satisfied,
and Lyapunov stability follows from
Lemma~\ref{lem:compositional-stability}.

For 2), combining the discretization error margin with
$\Sigma_i^d\succeq s_iI_{m_i}$ gives
\[
\Sigma_i
\succeq
\Sigma_i^d-\varepsilon_iI_{m_i}
\succeq
(s_i-\varepsilon_i)I_{m_i}.
\]
Moreover, the condition
$s_i-\sigma_{net}>\varepsilon_i$ yields
\[
s_i-\varepsilon_i>\sigma_{net},
\qquad \forall i\in\mathcal V.
\]
Hence, each continuous-time local ODP matrix admits
a scalar lower bound exceeding the network threshold.
The decentralized stability condition is therefore satisfied,
and Lyapunov stability follows from
Corollary~\ref{cor:decentralized-stability}.
\end{proof}

\proofheading{Proof of Theorem~\ref{thm:is-data-odp}}{app:proof-is-data}
\begin{proof}
Let $Z=[X^\top\;U^\top]^\top$.
Using $X_+=\bar{A}X+\bar{B}U$ and the definition of $M_3$, we obtain
\begin{align*}
X_+^\top PX_+-X^\top PX
&=Z^\top\bar{M}_1Z,\\
M_3
&=(U-\Sigma^d CX)^\top(CX_+-CX)\\
&=Z^\top\bar{M}_2Z.
\end{align*}
Define
\[
L:=\bar{M}_1-\frac{1}{2}(\bar{M}_2+\bar{M}_2^\top).
\]
The data-driven inequality~\eqref{eq:is-data-lmi}
is therefore equivalent to
\[
Z^\top LZ\preceq0.
\]
Condition~\eqref{eq:is-rank-condition} ensures that
$Z$ has full row rank. Hence, there exists a right inverse
$R$ satisfying $ZR=I_{n+m}$.
Premultiplying the above inequality by $R^\top$
and postmultiplying it by $R$ gives
\[
L=R^\top(Z^\top LZ)R\preceq0,
\]
which is precisely the discrete-time ODP
LMI~\eqref{eq:dt-odp-lmi}.
Conversely, $L\preceq0$ implies $Z^\top LZ\preceq0$
by congruence, recovering~\eqref{eq:is-data-lmi}.
Thus, the data-driven and model-based LMIs are equivalent.
The conclusion follows from
Theorem~\ref{thm:dt-odp-lmi}.
\end{proof}

\proofheading{Proof of Theorem~\ref{thm:io-data-odp}}{app:proof-io-data}
\begin{proof}
The extended-state data satisfy
\begin{equation}
\Xi=
\begin{bmatrix}
H_l(\{u_k\}_{k=0}^{N-2})\\
H_l(\{y_k\}_{k=0}^{N-2})
\end{bmatrix}.
\end{equation}
Let $Z_\Xi=[\Xi^\top U_\Xi^\top]^\top$.
The history representation gives
$\operatorname{rank}(Z_\Xi)
\le(l+1)m+\operatorname{rank}(O_l)$.
Thus, the lag-selection rank condition ensures
$\operatorname{rank}(O_l)=n$.
Since $Z_\Xi$ contains reachable state--input pairs
and has rank $n+(l+1)m$, its columns span all such pairs.

Using $\Xi_+=\bar{\widetilde A}\Xi+\bar{\widetilde B}U_\Xi$,
$Y_\Xi=\widetilde C\Xi$, and
$Y_{\Xi,+}=\widetilde C\Xi_+$,
the inequality~\eqref{eq:io-odp-lmi} can be rewritten as
\begin{equation}
Z_\Xi^\top
\left[
\bar{\widetilde M}_1-
\frac12(\bar{\widetilde M}_2+\bar{\widetilde M}_2^\top)
\right]
Z_\Xi\preceq0,
\end{equation}
where $\bar{\widetilde M}_1$ and $\bar{\widetilde M}_2$ are obtained
from~\eqref{eq:M1}--\eqref{eq:M2} by replacing
$(\bar{A},\bar{B},C)$ with
$(\bar{\widetilde A},\bar{\widetilde B},\widetilde C)$.

Denote the matrix in brackets by $\widetilde L$.
For any reachable state--input pair
$z_k=[\xi_k^\top\;u_k^\top]^\top$,
the spanning property ensures that
$z_k=Z_\Xi a_k$ for some coefficient vector $a_k$.
Consequently,
\[
z_k^\top\widetilde Lz_k
=a_k^\top Z_\Xi^\top\widetilde LZ_\Xi a_k
\leq0.
\]
Expanding this quadratic form shows that
$S_\xi(\xi)=\xi^\top P\xi$ satisfies
\[
S_\xi(\xi_{k+1})-S_\xi(\xi_k)
\leq
(u_k-\Sigma^d y_k)^\top(y_{k+1}-y_k)
\]
along every reachable trajectory of the extended realization.
To obtain a storage function on the original state space,
note that the lag condition gives a linear map $T$ such that
$x=T\xi$ for all $\xi\in\mathcal X_\xi^r$.
Since the original realization is minimal,
$T\mathcal X_\xi^r=\mathbb R^n$. Define
\[
S_x(x):=
\min_{\substack{\xi\in\mathcal X_\xi^r\\T\xi=x}}
\xi^\top P\xi .
\]
Because $P\succ0$ and the constraint map is surjective,
this minimum is a positive-definite quadratic form
$S_x(x)=x^\top P_xx$ with $P_x\succ0$.

For any state $x$ and input $u$, let $\xi_x$ denote
the minimizer and set
$\xi_+=\bar{\widetilde A}\xi_x+\bar{\widetilde B}u$.
Then $\xi_+\in\mathcal X_\xi^r$ and
$T\xi_+=x_+:=\bar{A}x+\bar{B}u$, with $y=Cx$ and $y_+=Cx_+$.
Consequently,
\[
\begin{aligned}
S_x(x_+)-S_x(x)
&\le S_\xi(\xi_+)-S_\xi(\xi_x)\\
&\le (u-\Sigma^d y)^\top(y_+-y).
\end{aligned}
\]
Thus, the original system admits a positive-definite
quadratic storage function satisfying the discrete-time
ODP inequality, and therefore satisfies
$\mathrm{ODP}(\Sigma^d)$.
\end{proof}

\proofheading{Proof of Theorem~\ref{thm:riw}}{app:proof-riw}
\begin{proof}
Substituting the whitened matrices $\widehat{\mathcal X}=\mathcal X W$,
$\widehat{\mathcal U}=\mathcal U W$, etc.\ into \eqref{eq:whitened_lmi}
yields the algebraic identity $\widehat{\mathcal F}(P,\Sigma^d)
=W^\top\mathcal F(P,\Sigma^d)W$, which holds for any data. It thus suffices to show $\mathcal F(P,\Sigma^d)\preceq0 \Leftrightarrow
W^\top\mathcal F(P,\Sigma^d)W\preceq0$.

For either formulation, the aligned data are generated by an LTI realization of the state variable $\eta_k$, i.e.,
\[
    \eta_{k+1}=\bar{A}_\eta\eta_k+\bar{B}_\eta u_k,\qquad
    y_k=C_\eta\eta_k .
\]

With $\mathcal X=[I\ 0]Z$ and $\mathcal U=[0\ I]Z$, the LTI relations give $\mathcal X_+=[\bar{A}_\eta\ \bar{B}_\eta]Z$, $\mathcal G=C_\eta[I\ 0]Z$, and $\mathcal G_+=C_\eta[\bar{A}_\eta\ \bar{B}_\eta]Z$. Thus, every data matrix appearing in \eqref{eq:unified_data_lmi} is a linear image of $Z$.

Hence, the matrix in \eqref{eq:unified_data_lmi} can be factorized as
\[
    \mathcal F(P,\Sigma^d)
    =
    Z^\top \mathcal H(P,\Sigma^d) Z
\]
for some symmetric matrix $\mathcal H(P,\Sigma^d)$. Using $Z=Q_rS_rV_r^\top$, it follows that
$\mathcal F(P,\Sigma^d)
=V_rS_rQ_r^\top \mathcal H(P,\Sigma^d)Q_rS_rV_r^\top$.
Moreover,
\[
    W^\top\mathcal F(P,\Sigma^d)W
    =
    Q_r^\top\mathcal H(P,\Sigma^d)Q_r .
\]
Let the columns of $V_\perp$ form an orthonormal basis of the orthogonal complement of $\operatorname{span}(V_r)$. The factorization implies $V_\perp^\top\mathcal F(P,\Sigma^d)V_\perp=0$ and
$V_\perp^\top\mathcal F(P,\Sigma^d)V_r=0$. Therefore,
\[
    \mathcal F(P,\Sigma^d)\preceq0
    \quad\Longleftrightarrow\quad
    S_rQ_r^\top\mathcal H(P,\Sigma^d)Q_rS_r\preceq0 .
\]
Since $S_r\succ0$, this is further equivalent to
\[
    W^\top\mathcal F(P,\Sigma^d)W
    =
    Q_r^\top\mathcal H(P,\Sigma^d)Q_r
    \preceq0. \qedhere
\]
\end{proof}

\bibliographystyle{IEEEtran}
\bibliography{bib/references}
\vspace{10pt}

\end{document}